\documentclass[%
 reprint,
 amsmath,amssymb,
 prd,
]{revtex4-2}
\pdfoutput=1
\usepackage{graphicx}
\usepackage[utf8]{inputenc}
\usepackage{lineno}

\begin{document}

\title{Insensitivity of the complexity rate of change to the conformal anomaly and Lloyd’s bound as a possible renormalization condition}

\author{Daniel \'Avila}
 \email{davhdz06@ciencias.unam.mx}
\author{C\'esar D\'iaz}
 \email{cadh@ciencias.unam.mx}
\author{Leonardo Pati\~no}
 \email{leopj@ciencias.unam.mx}
\affiliation{Departamento de F\'isica, Facultad de Ciencias, Universidad Nacional Aut\'onoma de M\'exico, \\  A.P. 70-542, M\'exico D.F. 04510, Mexico}

\author{Yaithd D. Olivas,}
 \email{yaithd.olivas@correo.nucleares.unam.mx}
\affiliation{Departamento de Física de Altas Energias, Instituto de Ciencias Nucleares,
Universidad Nacional Autónoma de México Apartado Postal 70-543, CDMX 04510, México}

\begin{abstract}
We determine the effect on the computational complexity of a conformal anomaly using the Complexity=Action prescription of the gauge/gravity correspondence. To allow the involvement of said anomaly, we extend previous studies to include arbitrary values for the anisotropic parameter and the magnetic field respectively on the Mateos-Trancanelli and the D'Hoker-Kraus holographic models. Our main result is that the rate of change of the computational complexity is independent of the conformal anomaly in both cases. In addition, this allows us to also show that, if so desired, the saturation of Lloyd's bound at infinite time can be used as a renormalization condition.
\end{abstract}


\maketitle
\section{Introduction}
\label{Intro}

In recent years, the AdS/CFT correspondence \cite{Maldacena:1997re} has served as a bridge that connects concepts of quantum information theory and gravity. The holographic dictionary \cite{Gubser:1998bc,Witten:1998qj} typically relates bulk and boundary physical quantities, such as the entanglement entropy, whose holographic dual is the area of an extremal surface in the bulk of the spacetime \cite{Ryu:2006bv,Hubeny:2007xt}, or the entanglement of purification, which is dual to the area of the minimal cross-section of the entanglement wedge \cite{Takayanagi:2017knl}. However, there is now evidence that the black hole interior also encodes important information about the dual gauge theory, not only concerning information theory \cite{Penington:2019npb,Almheiri:2019hni,Almheiri:2019yqk}, but other systems as well, such as superconductors \cite{Hartnoll:2020fhc}. An important step in this direction was the conjecture, based on studies of the time evolution of the entanglement entropy \cite{Hartman:2013qma}, that the computational complexity of the boundary state is also affected by the interior of the black hole. 

There are two main proposals for this connection, namely the Complexity=Volume (CV) \cite{Susskind:2014rva,Stanford:2014jda} and the Complexity=Action (CA) \cite{Brown:2015bva,Brown:2015lvg} conjectures. In the CV conjecture, the complexity is dual to the volume of a certain extremal region in the bulk \cite{Alishahiha:2015rta}. However, the precise prescription includes some unsatisfactory features like the introduction of an arbitrary scale and the necessity for a special foliation of spacetime. Tackling these issues, the CA conjecture was proposed as a refinement in which complexity is dual to the action of the gravitational theory evaluated in a region known as the Wheeler-DeWitt (WDW) patch \cite{Brown:2015bva,Brown:2015lvg}, with the precise relation being
\begin{equation}
C=\frac{S_{WDW}}{\pi}.
\label{CA}
\end{equation}
The WDW patch is the region enclosed by past and future light sheets sent into the bulk from a constant time slice on the boundary, where the complexity is meant to be evaluated. In other words, the WDW patch is the domain of dependence of any Cauchy surface in the bulk which asymptotically approaches said time slice at the boundary.

The gravity setup considered in \cite{Brown:2015bva,Brown:2015lvg} was a two-sided black hole geometry. From the point of view of the gauge theory, this is dual to a thermofield double (TFD) state which, denoting the two asymptotic regions as the left (L) and right (R) boundaries, can be written schematically as
\begin{equation}
|TFD\rangle=\frac{1}{Z^{\frac{1}{2}}}\sum_{n}e^{-\frac{\beta E_{n}}{2}}e^{-E_{n}(t_{L}+t_{R})}|E_{n}\rangle_{L}|E_{n}\rangle_{R},
\end{equation}
where $\beta$ is the inverse of the temperature. It is important to note that the TFD state is invariant under time evolution with the Hamiltonian $H=H_{L}-H_{R}$, which means that the state is invariant under the transformations $t_{L}\rightarrow t_{L}+\Delta t$ and $t_{R}\rightarrow t_{R}-\Delta t$. This implies that  the TFD state only depends on time through the combination $\tau=t_{L}+t_{R}$.

The late time behavior of the rate of change of the complexity of certain black holes was studied in \cite{Brown:2015bva,Brown:2015lvg} using both the CV and the CA conjectures. In the case of a planar five-dimensional AdS-Schwarzschild black hole the CV late-time calculation leads to
\begin{equation}
\lim_{t\rightarrow\infty}\frac{dC}{d\tau}=4\pi E,
\end{equation}
while in the case of CA the behavior is
\begin{equation}
\lim_{t\rightarrow\infty}\frac{dC}{d\tau}=\frac{2E}{\pi},
\end{equation}
 which is consistent with Lloyd’s bound \cite{LloydBound} on the rate of computation by a system with energy $E$ 
 \begin{equation}
 \frac{dC}{d\tau}\leq\frac{2E}{\pi}.
 \label{Lloyd}
 \end{equation}

It was later shown \cite{Lehner:2016vdi} that the correct evaluation of the gravitational action over the WDW patch requires the addition of boundary terms $S_{surf}$ resulting from the codimension-one null hypersurfaces that delimit part of it, and of joint terms $S_{joint}$ arising from the codimension-two intersections of the light sheets with the boundary and the singularity. It was also observed that there is an ambiguity in $S_{surf}$ coming from the normalization of normal vectors to the null segments of the boundary, and that it is necessary to add a counterterm $S_{null}$ in order to remove it. Hence the full action can be written schematically as
\begin{equation}
S=S_{bulk}+S_{surf}+S_{joint}+S_{null}.
\label{Action}
\end{equation}

The complete action \eqref{Action} made it possible in \cite{Carmi:2017jqz} to study the behavior at all times for the rate of change of the complexity. It was shown that, for the black holes that they consider, the CV prescription saturates Lloyd's bound \eqref{Lloyd} from below at late times, while the CA prescription saturates it from above. In other words, it was shown that the CA prescription violates Lloyd's bound for any finite time, saturating it only at infinite time. This led to question the validity of Lloyd's bound in the holographic context, whether it is reasonable to expect it to hold, and under which conditions \cite{Cottrell:2017ayj}. For instance, it was found in \cite{Couch:2017yil} that Lloyd's bound is violated for nonconmutative SYM theories, and the same was shown in \cite{Swingle:2017zcd,Alishahiha:2018tep} for the case of Lifshitz and hyperscaling violating theories, even at late times. Other examples of violations of Lloyd's bound can be found in \cite{Alishahiha:2017hwg,Mahapatra:2018gig,Chen:2020qty,Babaei-Aghbolagh:2020vsz}. There is also the question of exactly what kind of complexity is dual to the holographic complexity \cite{Yang:2020tna}. For example, Nielsen’s circuit complexity \cite{Nielsen} for a charged thermofield double state was compared to the holographic result using CA in \cite{Doroudiani:2019llj}, obtaining a different saturation time for Lloyd's bound depending on the approach.

Recently, the effect that a spatial anisotropy in the gauge theory can have on the rate of change of the complexity was studied in \cite{HosseiniMansoori:2018gdu} using certain limits of the Cheng-Ge-Sin (CGS) model, the Mateos-Trancanelli (MT) anisotropic model \cite{Mateos:2011ix,Mateos:2011tv}, and the D'Hoker-Krauss (DK) magnetic model \cite{DHoker:2009mmn}. The results show that at late times Lloyd's bound is saturated from above for small anisotropies in the case of the MT model and for an infinite magnetic field intensity in the case of the DK model. However, both models feature a conformal anomaly for any non-vanishing $a$ in the case of the MT model and any non-vanishing $b$ in the case of the DK model. The conformal anomaly (also known as the Weyl or scale anomaly) was first studied in the holographic context in \cite{Henningson:1998gx}, where it was shown that the Lagrangian of the theory is no longer invariant under rescalings of the holographic radial coordinate when a logarithmic divergence appears in the holographic renormalization procedure \cite{deHaro:2000vlm,Bianchi:2001kw,Skenderis:2002wp,Papadimitriou:2005ii}. It was shown in \cite{ Bianchi:2001de} that this is related to the appearance of a reference energy scale, that is independent of those in the classic counterpart. The existence of this parameter that is not present in the classical theories manifests itself by making some physical observables, such as the expectation value of the stress energy tensor of the gauge theory, to have anomalous behavior under scale transformations. The non-trivial manner in which the conformal anomaly present \cite{Henningson:1998gx,deHaro:2000vlm,Papadimitriou:2005ii,Bianchi:2001de} in both the MT and DK models can affect the physics of the theory was not studied as part of the limits above, as it becomes a subdominant contribution of order $(a/T)^4$ \cite{Mateos:2011ix,Mateos:2011tv} for the first of these models, while the second undergoes a dimensional reduction \cite{Martinez-y-Romero:2017awl} in the $b/T^2\rightarrow\infty$ limit that renders the four-dimensional conformal anomaly inapplicable.

The main objective of this work is to determine the effect that the conformal anomaly has on the complexity of the state, and in particular whether or not it has an impact on Lloyd's bound. In order to achieve this, we extend the study presented in \cite{HosseiniMansoori:2018gdu} to include arbitrary anisotropies and magnetic field intensities, finding as our main result that, even if the complexity itself is affected by the conformal anomaly, its rate of change is not, making it independent of the renormalization scale. Given that the energy of the state depends on this scale, we also show that the saturation of Lloyd's bound at infinite time can be used, if so desired, as a renormalization condition and use a scheme related constant that appears in the construction to preserve it at all values of $a/T$ in the MT model and of $b/T^2$ for the DK case. 

The manuscript is organized as follows. In Sec. \ref{GravitySetup} we review the construction of both the MT and DK models, explaining the role of the conformal anomaly in both cases. The details about the construction of the interior solutions can be consulted in the Appendix. In Sec. \ref{WDWPatch} we explain the construction of the WDW patch for both models, and in Sec. \ref{Complexity} we compute the complexity of the TFD state, along with its rate of change, using the CA prescription. We close by discussing our results in Sec. \ref{Discussion}.

\section{Gravity setup}
\label{GravitySetup}
\subsection{Anisotropic black branes}
The Mateos-Trancanelli (MT) anisotropic model \cite{Mateos:2011ix,Mateos:2011tv} is a ten-dimensional family of solutions to SUGRA IIB. However, for our purposes it suffices to consider its reduction to five dimensions, which action in the Einstein frame is given by\footnote{Our curvature convention for both models is ${R^{\alpha}}_{\beta\mu\nu}=\partial_{\mu}{\Gamma^{\alpha}}_{\beta\nu}+{\Gamma^{\alpha}}_{\mu\sigma}{\Gamma^{\sigma}}_{\beta\nu}-(\mu\leftrightarrow\nu)$ and $R_{\mu\nu}={R^{\sigma}}_{\mu\sigma\nu}$}
\begin{eqnarray}
&& S  =\frac{1}{16\pi G_{5}}\int dx^{5}\sqrt{-g}\left(R+\frac{12}{L^{2}}-\frac{1}{2}(\partial\phi)^{2}-\frac{1}{2}e^{2\phi}(\partial\chi)^{2}\right) \cr
&& \qquad +\frac{1}{16\pi G_{5}}\int dx^{4}\sqrt{-\gamma}2K,
\label{AnisotropicAction}
\end{eqnarray}
where $\phi$, $\chi$ and $g_{\mu\nu}$ are the dilaton, the axion and metric fields respectively. $G_{5}$ is the five-dimensional Newton constant and $L$ is the $AdS_{5}$ radius, which we will set to unity without loss of generality\footnote{Taking $L=1$ implies that $G_{5}=\pi/2N_{c}^{2}$ with $N_{c}$ the number of color degrees of freedom in the dual theory.}. The second integral is the York-Gibbons-Hawking (YGH) surface term, in which $\gamma_{ij}$ is the induced metric at the boundary and $K$ its extrinsic curvature. 

Every member of the MT family of solutions is part of the ansatz 
\begin{eqnarray}
&& ds^{2}=e^{-\frac{1}{2}\phi}r^{2}\left(\frac{dr^{2}}{r^{4}\mathcal{F}}-\mathcal{F}\mathcal{B}dt^{2}+dx^{2}+dy^{2}+e^{-\phi}dz^{2}\right),\cr
&& \chi=a\,z, \qquad \phi=\phi(r).
\label{AnisotropicAnsatz}
\end{eqnarray}
Here $r$ is the $AdS_{5}$ radial coordinate, in terms of which the boundary is located at $r\rightarrow\infty$. All these backgrounds feature a horizon located at $r=r_{h}$, where the metric function $\mathcal{F}(r)$ vanishes. The parameter $a$ measures the degree of anisotropy in the geometry, and its related to a density of D7-branes that are smeared in the geometry from the ten-dimensional perspective (see \cite{Mateos:2011ix,Mateos:2011tv} for more details). Meanwhile, the temperature of each member of the family is given by
\begin{eqnarray}
T=\frac{e^{-\frac{1}{2}\phi(r_{h})}\sqrt{\mathcal{B}(r_{h})}(16 r_{h}^{2}+a^{2}e^{\frac{7}{2}\phi(r_{h})})}{16\pi r_{h}}.
\end{eqnarray}

The solutions asymptote $AdS_{5}$ for any value of $a$ and $T$ when the limit $r\rightarrow\infty$ is taken. In this region the metric functions and dilaton are given by
\begin{eqnarray}
&& \mathcal{F}(r)=1+\frac{11 a^{2}}{24 r^{2}}+\frac{1}{r^4}\left(f_{4}-\frac{7}{12}a^{4}\log{r}\right) +O\left(\frac{1}{r^{6}}\right),
\cr
&& \mathcal{B}(r)=1-\frac{11 a^{2}}{24 r^{2}}+\frac{1}{r^4}\left(b_{4}+\frac{7}{12}a^{4}\log{r}\right) +O\left(\frac{1}{r^{6}}\right),
\cr
&& \phi(r)=-\frac{a^{2}}{4 r^{2}}+\frac{1}{r^4}\left(\frac{121 a^{4}}{4032}+\frac{2 b_{4}
}{7}+\frac{a^{4}}{6}\log{r}\right) \cr 
&& \qquad +O\left(\frac{1}{r^{6}}\right).
\label{AnisotropicBoundary}
\end{eqnarray}
The two coefficients $b_{4}$ and $f_{4}$ in \eqref{AnisotropicBoundary} are not determined by the equations of motion, but can be read once a numerical solution is known. It is because of this that both are functions of the anisotropy $a$ and temperature $T$ of the solution, as well of the renormalization scale $\mu$ that will be discussed below.

The metric functions and the dilaton are known analytically in the limits of high and low temperature \cite{Mateos:2011tv}, while for intermediate regimes one has to resort to numerics to solve the equations of motion coming from \eqref{AnisotropicAction}. The explicit integration procedure is explained in detail in \cite{Mateos:2011tv}, where the solutions were computed outside the horizon. However, in order to apply the CA prescription we need to know the solutions inside the horizon, as the WDW patch extends in this region. We describe the integration procedure needed for this in App. \ref{AppAnisotropic}.

From the above discussion we can see that every member of the family of solutions is characterized by the value of its anisotropy $a$ and temperature $T$, which seem to be the only two parameters with dimensions of length, inviting us to label each solution by the dimensionless ratio $a/T$. However, it turns out that not all dimensionless physical observables are functions of this ratio alone, indicating the presence of a conformal anomaly \cite{Henningson:1998gx} in the dual to any of the backgrounds above with $a\neq 0$. This anomaly can be explicitly exhibit by taking the trace of the stress-energy tensor in the corresponding gauge theory, obtained from the variation with respect to the boundary metric of the on-shell evaluations of the action \eqref{AnisotropicAction}. As it is usually the case in holography, the result of said evaluation diverges when the integration is taken all the way up to the boundary. To deal with this issue, the subtraction of the divergent behavior has to be done by adding covariant boundary terms to the action, in a process known as holographic renormalization \cite{Skenderis:2002wp,Bianchi:2001kw}. 

The counterterm action for the MT model is given by \cite{Mateos:2011tv}
\begin{equation}
\begin{split}
S_{ct}=&\frac{1}{8\pi G_{5}}\left(\int d^{4}x\sqrt{-\gamma}\left(3-\frac{1}{8}e^{2\phi}\partial_{i}\chi\partial^{i}\chi\right) \right. \\& \left. +\log{r}\int d^{4}x\sqrt{-\gamma}\frac{e^{4\phi}}{12}(\partial_{i}\chi\partial^{i}\chi)^{2}\right.\\& \left.+\frac{1}{4}(C_{sch}-1)\int d^{4}x\sqrt{-\gamma}\frac{e^{4\phi}}{12}(\partial_{i}\chi\partial^{i}\chi)^{2}\right),
\label{AnisotropicCounterterms}
\end{split}
\end{equation}
where all the integrals are performed over a constant $r$ surface and the limit $r\rightarrow\infty$ is meant to be taken. Also, we are denoting the boundary coordinates as $x^{i}$, and all the contractions are taken using the metric $\gamma_{ij}$ induced at the boundary. The first two terms are the minimum required to eliminate the divergences of the action \eqref{AnisotropicAction}, while the third gives a contribution that remains finite in the $r\rightarrow\infty$ limit. The freedom to add such a finite term is associated with the existence of the anisotropy related conformal anomaly discussed above, that has as a consequence the introduction of an independent arbitrary energy scale $\mu$. This is the reason why some dimensionless physical quantities, such as the energy density of the system $E/T^{4}$, do not depend only on the dimensionless ratio $a/T$, but on any two independent ratios that can be built from $a$, $T$ and $\mu$.

In \eqref{AnisotropicCounterterms} we have fixed $\mu=L=1$ in the argument of the logarithm of $r$. Any change in this renormalization scale can be absorbed into the finite term by choosing a different value for the $C_{sch}$ coefficient. We can therefore use this coefficient to reverse the change that would be introduced in physical quantities to keep a particular renormalization scheme fixed as the energy scale $\mu$ is modified, attaining as a final consequence the scheme independence of relevant physical results. It is in this sense that $C_{sch}$ is a scheme-dependent quantity (see \cite{Ecker:2017fyh} for a more detailed discussion).

\begin{figure}[ht!]
 \centering
 \includegraphics[width=0.4\textwidth]{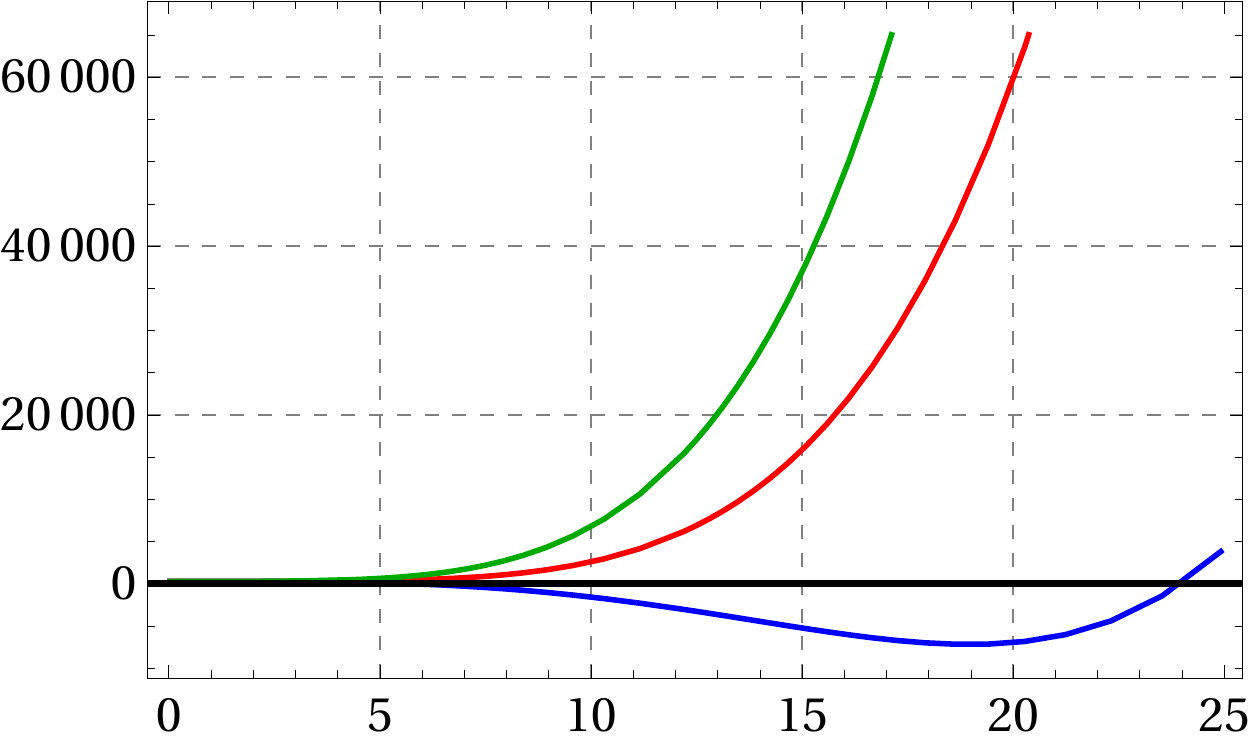}
 \put(-0,-10){\Large $\frac{a}{T}$}
 \put(-225,120){\Large $\frac{E}{\mathcal{N}T^{4}}$}
\caption{\small Energy density for the MT model as a function of $a/T$. Blue, red and green curves (bottom to top) correspond to $C_{sch}=\{-10,0,10\}$ respectively.}
\label{AnisotropicEnergy}
\end{figure}

With the counterterms in hand, the stress-energy tensor of the system can be computed from the renormalized action
\begin{eqnarray}
S_{ren}=S+S_{ct},
\end{eqnarray}
by taking its variation with respect to the boundary metric and evaluating the result at the boundary using the expansions \eqref{AnisotropicBoundary}. This procedure gives \cite{Mateos:2011ix,Mateos:2011tv}
\begin{equation}
\langle T_{ij}\rangle=\text{diag}(E,P^{\perp},P^{\perp},P^{\parallel}),
\end{equation}
where $E$ is the energy of the state, while $P^{\perp}$ and $P^{\parallel}$ are respectively the pressures along directions perpendicular and parallel to the anisotropic direction. These quantities are given by
\begin{eqnarray}
&& E=\mathcal{N}\left(-3f_{4}-\frac{23}{7}b_{4}+\frac{2777}{4032}a^{4}+\frac{C_{sch}}{24}a^{4}\right),
\cr
&& P^{\perp}=\mathcal{N}\left(-f_{4}-\frac{5}{7}b_{4}+\frac{611}{4032}a^{4}-\frac{C_{sch}}{24}a^{4}\right),
\cr
&& P^{\parallel}=\mathcal{N}\left(-f_{4}-\frac{13}{7}b_{4}+\frac{2227}{4032}a^{4}+\frac{C_{sch}}{8}a^{4}\right),
\label{AnisotropicTmunu}
\end{eqnarray}
where the explicit dependence on the three parameters $a$, $T$, and $\mu$ has to be supplemented by the one implicit in $f_{4}$ and $b_{4}$ that was previously discussed, and the normalization constant $\mathcal{N}$ is given by
\begin{eqnarray}
\mathcal{N}=\frac{V_{x}}{16\pi G_{5}},
\label{Ngarabato}
\end{eqnarray}
with $V_{x}$ the spatial volume of the boundary region under consideration. From this expressions we can see that the trace of the stress-energy tensor is
\begin{equation}
\langle {T^{i}}_{i}\rangle=\frac{\mathcal{N}}{6}a^{4},
\end{equation}
which does not vanish for any $a\neq 0$, showing the existence of the conformal anomaly. In Fig. (\ref{AnisotropicEnergy}) we show the quantity $E/\mathcal{N}T^{4}$ in the MT model as a function of $a/T$ for three different values of $C_{sch}$, as we do not have any renormalization condition to fix the scheme at this stage.

\subsection{Magnetic black branes}
\label{MagneticBackground}
The D'Hoker-Kraus (DK) model \cite{DHoker:2009mmn} is a family of solutions to the five-dimensional gauged supergravity which bosonic part of the action is given by
\begin{equation}
\begin{split}
S=&\frac{1}{16\pi G_5}\left[\int d^5x\sqrt{-g}\left(R-F^{2}+\frac{12}{L^{2}}\right) \right. \\& +  \left.\frac{8}{3\sqrt{3}}\int A\wedge F\wedge F+\int dx^{4}\sqrt{-\gamma}2K\right],
\end{split}
\label{MagneticAction}
\end{equation}
where $F$ and $g_{\mu\nu}$ are respectively the Maxwell and metric fields, $G_{5}$ is the five-dimensional Newton constant, and $L$ is the $AdS_{5}$ radius, which we will set again to unity without loss of generality. The last integral is the York-Gibbons-Hawking (YGH) surface term, in which $\gamma_{ij}$ is the induced metric at the boundary and $K$ its extrinsic curvature. This theory is a consistent truncation of ten-dimensional SUGRA IIB \cite{Cvetic:1999xp} and the backgrounds that we are about to study were uplifted in \cite{Elinos:2021bmx} to turn them into configurations that solve the equations proper of the latter theory.

Every member of the DK family of solutions is part of the ansatz 
\begin{eqnarray}
&& ds^{2}=\frac{dr^{2}}{U(r)}-U(r)dt^{2}+V(r)(dx^{2}+dy^{2})+W(r)dz^{2},
\cr
&& F=b\,dx\wedge dy,
\label{MagneticAnsatz}
\end{eqnarray}
where $r$ is the $AdS_{5}$ radial coordinate, in terms of which the boundary is located at $r\rightarrow\infty$.  All these backgrounds feature a horizon located at $r_{h}$ where the metric function $U(r)$ vanishes. The magnetic field intensity $b$ matches the one in the dual gauge theory given that the metric asymptotes precisely $AdS_{5}$ at the boundary. From the ten-dimensional perspective this Maxwell field is interpreted as an infinitesimal rotation in the compact part of the geometry (see \cite{Cvetic:1999xp,Avila:2020ved} for additional details). Meanwhile, the temperature of each solution is given by
\begin{equation}
T=\frac{3 r_{h}}{2\pi}.
\label{MagneticTemperature}
\end{equation}
Thus every member of the family is characterized by the values of its magnetic field intensity $b$ and temperature $T$, which at first sight seem to be the only two parameters with dimensions of length, suggesting the labelling of each solution by the dimensionless ratio $b/T^{2}$.

The only known analytical members of this family are the Schwarzschild-AdS black brane for $b/T^{2}=0$ and BTZ$\times\mathbb{R}^{2}$ for precisely $b/T^{2}=\infty$. For any intermediate values it is necessary to resort to numerical methods to solve the equations coming from \eqref{MagneticAction}. The explicit integration procedure that we follow is explained in detail in \cite{Arean:2016het} for the exterior solutions, and in \cite{Avila:2018sqf} for the interior solutions. We review both in App. \ref{AppMagnetic}. The backgrounds are constructed so that for any value of $b/T^{2}$ the geometry asymptotes $AdS_{5}$ when the limit $r\rightarrow\infty$ is taken. In this region the functions in the line element \eqref{MagneticAnsatz} are given by
\begin{eqnarray}
&& U(r)=r^{2}+u_{1}r+\frac{u_{1}^{2}}{4}+\frac{1}{r^{2}}\left(u_{4}-\frac{2}{3}b^{2}\log{r}\right)+\mathcal{O}\left(\frac{1}{r^{4}}\right),
\cr
&& \begin{split} V(r)=&r^{2}+u_{1}r+\frac{u_{1}^{2}}{4}+\frac{1}{r^{2}}\left(-\frac{1}{2}w_{4}+\frac{1}{3}b^{2}\log{r}\right)\\&+\mathcal{O}\left(\frac{1}{r^{4}}\right),\end{split}
\cr
&& \begin{split} W(r)=&r^{2}+u_{1}r+\frac{u_{1}^{2}}{4}+\frac{1}{r^{2}}\left(w_{4}-\frac{2}{3}b^{2}\log{r}\right)\\&+\mathcal{O}\left(\frac{1}{r^{4}}\right).\end{split}
\label{MagneticBoundary}
\end{eqnarray}
The three coefficients $u_{1}$, $w_{4}$, and $u_{4}$ are not determined by the equations of motion, but can be read from each numerical solution. Consequently all three coefficients are functions of the magnetic field $b$ and the temperature $T$ of the background, through dimensionless combinations involving the renormalization scale $\mu$ that will be introduced below.

While it would seem like any member of the family of solutions can be characterized by the dimensionless ratio $b/T^{2}$, not every physical observable is a function of this ratio solely. Just like the MT model, the dual of any solution with non-vanishing magnetic field $b\neq 0$ features a conformal anomaly. In order to obtain the stress energy tensor required to support the latter claims, it is necessary to compute the on-shell action \eqref{MagneticAction} and take its variation with respect to the boundary metric. The result of this once again diverges when the integration is taken all the way up to the boundary. 

The counterterm action for the DK model is given by \cite{Fuini:2015hba,Endrodi:2018ikq}
\begin{equation}
S_{ct}=-\frac{1}{16\pi G_{5}}\int\sqrt{-\gamma}\left(6-F^{ij}F_{ij}\log{r}+C_{sch}F^{ij}F_{ij}\right),
\label{MagneticCounters}
\end{equation}
where the integral is performed over a constant $r$ surface and the limit $r\rightarrow\infty$ is meant to be taken. The first two terms are the minimum required to remove the divergences in the action \eqref{AnisotropicAction}, while the third gives a finite contribution in the limit $r\rightarrow\infty$. As discussed previously for the MT model, the freedom to add a finite term to the action is related to the existence of a conformal anomaly in the theory, which introduces an arbitrary energy scale $\mu$. We fixed this scale to $\mu=L=1$ in \eqref{MagneticCounters}. 

\begin{figure}[ht!]
 \centering
 \includegraphics[width=0.4\textwidth]{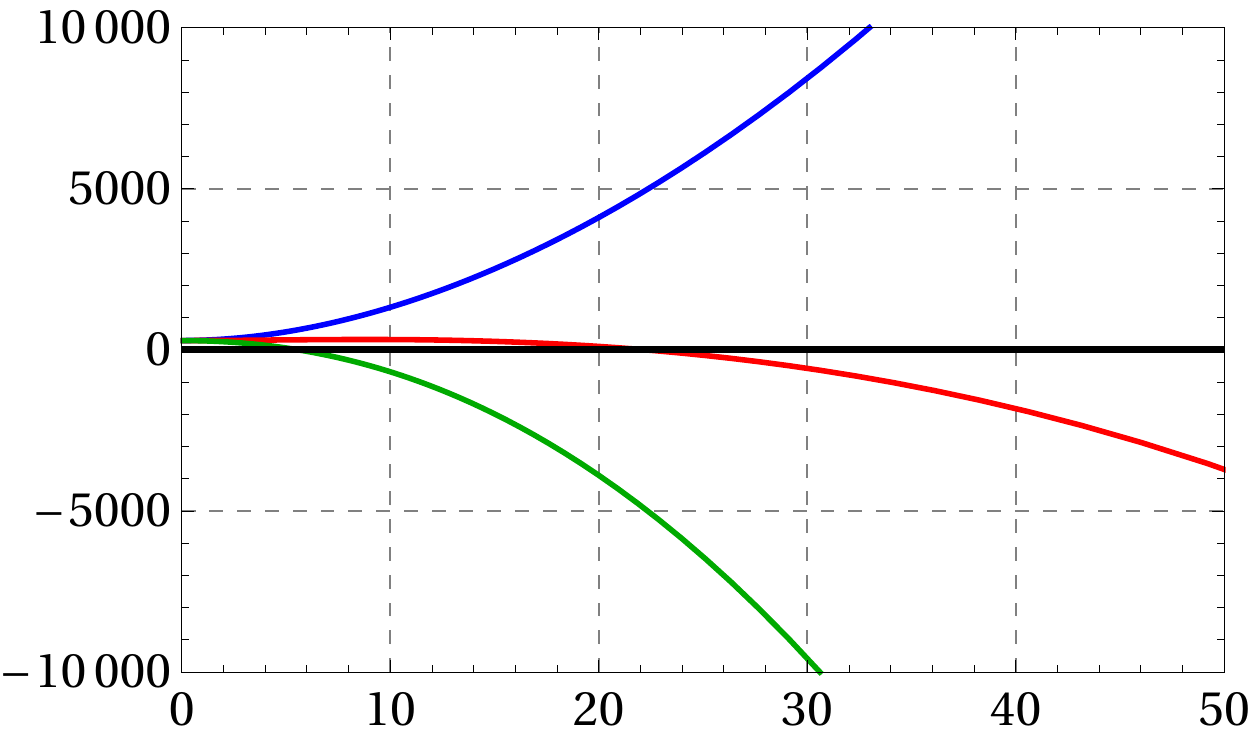}
 \put(-0,-10){\Large $\frac{b}{T^{2}}$}
 \put(-220,125){\Large $\frac{E}{\mathcal{N}T^{4}}$}
\caption{\small Energy density for the DK model as a function of $b/T^{2}$. Blue, red and green curves (top to bottom) correspond to $C_{sch}=\{-5,0,5\}$ respectively.}
\label{MagneticEnergy}
\end{figure}

The stress-energy tensor of the system can be computed from the renormalized action
\begin{eqnarray}
S_{ren}=S+S_{ct},
\end{eqnarray}
by taking its variation with respect to the boundary metric and evaluating the result at the boundary using the expansions \eqref{MagneticBoundary}. The result of doing this is \cite{Fuini:2015hba,Endrodi:2018ikq}
\begin{equation}
\langle T_{ij}\rangle=\text{diag}(E,P^{\perp},P^{\perp},P^{\parallel}),
\end{equation}
where $E$ is the energy density of the state in the gauge theory, while $P^{\perp}$ and $P^{\parallel}$ are respectively the pressures along directions perpendicular and parallel to the magnetic field. This quantities are explicitly given by
\begin{eqnarray}
&& E=\mathcal{N}\left(-3u_{4}-2C_{sch}b^{2}\right),
\cr
&& P^{\perp}=\mathcal{N}\left(-u_{4}-2w_{4}-b^{2}(1+2C_{sch})\right),
\cr
&& P^{\parallel}=\mathcal{N}\left(-u_{4}+4w_{4}+2C_{sch}b^{2}\right),
\label{MagneticTmunu}
\end{eqnarray}
where $u_{4}$ and $w_{4}$ depend on $b$, $T$, and $\mu$ as previously discussed, and the normalization constant $\mathcal{N}$ is the same as in \eqref{Ngarabato}. Using this expressions we can take the trace of the stress-energy tensor to find
\begin{equation}
\langle {T^{i}}_{i}\rangle=-2\mathcal{N}b^{2},
\end{equation}
which is non-vanishing for any $b\neq 0$, showing the existence of the conformal anomaly. In Fig. (\ref{MagneticEnergy}) we show the quantity $E/\mathcal{N}T^{4}$ in the DK model as a function of $b/T^{2}$ for three different values of $C_{sch}$.

\subsection{Penrose Diagram}

In this section we will derive the Penrose diagram for both the MT and DK model. The metric anzats for both are of the form
\begin{equation}
\begin{split}
ds^{2}=&g_{rr}(r)dr^{2}+g_{tt}(r)dt^{2}+g_{xx}(r)dx^{2}\\&+g_{yy}(r)dy^{2}+g_{zz}(r)dz^{2},
\label{PenroseMetric}
\end{split}
\end{equation}
where we are including the gauge theory spatial directions, even if they are not relevant for the derivation of the Penrose diagram. That being said, it is important to know that the metric functions $g_{rr}$ and $g_{tt}$ strongly depend on the source of the anisotropy, either $a$ or $b$. This two functions are such that $g_{tt}$ has a zero at $r_{h}$, while $g_{rr}$ has a simple pole there\footnote{Of course, the MT and DK models satisfy this conditions for any anisotropy and magnetic field.}. 

In order to construct the Penrose diagram for the metric \eqref{PenroseMetric} we first change to the tortoise coordinate $r_{\star}$, which is given by the solution to the equation
\begin{equation}
\frac{dr_{\star}}{dr}=\text{sign}(g_{rr})\sqrt{\left|\frac{g_{rr}}{g_{tt}}\right|},
\label{Tortoise}
\end{equation}
that satisfies the boundary condition $r_{\star}(\infty)=0$. Note that this coordinate automatically satisfies $r_{\star}\rightarrow\log(r-r_{h})$ as $r\rightarrow r_{h}$ because of the simple pole of $g_{rr}$ and the zero of $g_{tt}$ at $r_{h}$. Next we transform to the Kruskal-Szekeres coordinates, given by

\begin{eqnarray}
&& \mathcal{U}=+e^{-2\pi T(t-r_{\star})}, \quad \mathcal{V}=-e^{2\pi T(t+r_{\star})} \quad \text{Left exterior}
\cr
&& \mathcal{U}=-e^{-2\pi T(t-r_{\star})}, \quad \mathcal{V}=+e^{2\pi T(t+r_{\star})} \quad \text{Right exterior}
\cr
&& \mathcal{U}=+e^{-2\pi T(t-r_{\star})}, \quad \mathcal{V}=+e^{2\pi T(t+r_{\star})} \quad \text{Future interior}
\cr
&& \mathcal{U}=-e^{-2\pi T(t-r_{\star})}, \quad \mathcal{V}=-e^{2\pi T(t+r_{\star})} \quad \text{Past interior} \cr
&&
\end{eqnarray}
where $T$ is the temperature of the black hole. Finally, we change to the compact coordinates
\begin{equation}
X=\frac{\arctan \mathcal{V}-\arctan \mathcal{U}}{2}, \quad Y=\frac{\arctan \mathcal{V}+\arctan \mathcal{U}}{2},
\end{equation}
which leave the metric as
\begin{equation}
ds^{2}=\pm \frac{g_{tt}(X,Y)e^{-4\pi T r_{\star}}}{4(\pi T)^{4}}(1+\mathcal{V}^{2})(1+\mathcal{U}^{2})(-dY^{2}+dX^{2}).
\end{equation}
We present the Penrose diagram for both models, at different values of $a/T$ and $b/T^{2}$, in Fig. (\ref{AnisotropicWDW}) and (\ref{MagneticWDW}) respectively. Note that in the case of the DK model in Fig. (\ref{MagneticWDW}) the position of the singularity changes as $b/T^{2}$ increases, as explained in App. \ref{AppMagnetic}. For the analysis ahead it is also relevant to notice that the hypersurface $t=0$ corresponds to a horizontal line in the middle of the Penrose diagram of either model.

\section{WDW Patch}
\label{WDWPatch}

The WDW patch is the region enclosed by past and future light sheets extending into the bulk from a constant time slice on the boundary, where the complexity is meant to be evaluated. Because the TFD state only depends on time through the combination $t_{L}+t_{R}$, without loss of generality we will adopt the convention\footnote{Note that this means that $\tau=2t_{0}$, with $\tau$ the time variable relevant for the evolution of the TFD state mentioned in the introduction.} $t_{L}=t_{R}=t_{0}$. The direct evaluation of the action on the WDW patch is divergent, as it extends to the boundary at $r=\infty$ and to the singularity at $r=r_{s}$. To avoid this, we regularize it by introducing the cutoffs $r_{\max}$ near the boundary and $r_{\min}$ near the singularity. At the end of the calculation we will remove this regulators by taking the limits $r_{\max}\rightarrow\infty$ and $r_{\min}\rightarrow r_{s}$.

For a given boundary time $t_{0}$, the light sheets that delimit the WDW patch are given by
\begin{eqnarray}
&& r_{\star}(r)+t=t_{0} \qquad \text{right future light sheet},
\cr
&& r_{\star}(r)-t=-t_{0} \qquad \text{right past light sheet},
\cr
&& r_{\star}(r)-t=t_{0} \qquad \text{left future light sheet},
\cr
&& r_{\star}(r)+t=-t_{0} \qquad \text{left past light sheet}.
\label{LightSheets}
\end{eqnarray}
The names come from which of the singularities the light sheet reaches for $t_0=0$. Note that the difference in signs between the left and right regions comes from the fact that in the left region the flow of time is reversed, but we are taking $t_{0}=t_{L}=t_{R}$ by definition.

At early times, close to $t_0=0$, the WDW patch intersects both the past and future singularities which, as we will see in the following, makes its volume constant for some period of time $0\leq t_0\leq t_{c}$. However, for later times $t_0>t_{c}$, the WDW patch no longer intersects the past singularity and its volume reduces as time passes. This $t_{c}$ is known as the critical time, and is given by
\begin{equation}
t_{c}=-r_{\star}(r_{s}),
\label{tc}
\end{equation}
where we used the equations for the past light sheets in \eqref{LightSheets} to find the $t_0$ for which they intersect exactly at the singularity. Thus, for $0\leq t_{0}\leq t_{c}$ and any $(x,y,z)$ we can naturally divide the WDW patch in four regions
\begin{equation}
\begin{array}{lll}
I &=\{(t,r)|r\in[r_{min},r_{h}],\,t\in[-t_{0}+r_{\star}(r),t_{0}-r_{\star}(r)]\},
\\
II &=\{(t,r)|r\in[r_{h},r_{max}],\,t\in[t_{0}+r_{\star}(r),t_{0}-r_{\star}(r)]\},
\\
III &=\{(t,r)|r\in[r_{min},r_{h}],\,t\in[t_{0}+r_{\star}(r)),-t_{0}-r_{\star}(r)]\},
\\
IV &=\{(t,r)|r\in[r_{h},r_{max}],\,t\in[-t_{0}+r_{\star}(r),-t_{0}-r_{\star}(r)]\}.
\end{array}
\label{WDWRegionsEarly}
\end{equation}
We ilustrate this regions on the Penrose diagram in Fig. (\ref{AnisotropicWDW}) (a) for the MT model and in Fig. (\ref{MagneticWDW}) (b) for the DK model.

For later times the WDW patch does not reach the past singularity and instead ends on a minimal radius $r_{m}$ given by the intersection of the left and right past light sheets. Thus it is defined implicity by the solution of the equation
\begin{equation}
r_{\star}(r_{m})=-t_{0},
\label{rm}
\end{equation}
as a function of the fixed boundary time $t_{0}$. It will be useful to have an expression for the derivate of $r_{m}$ with respect to $t_{0}$. We can manipulate the derivative of the previous expression to get
\begin{equation}
\frac{dr_{m}}{dt_{0}}=-\left.\text{sign}(g_{rr})\sqrt{\left|\frac{g_{tt}}{g_{rr}}\right|}\right|_{r=r_{m}},
\label{rmderivative}
\end{equation}
where we have used \eqref{Tortoise}. The critical time for Schwarzschild-$AdS_{5}$ can be computed analitically, and the result is
\begin{equation}
t_{c}=\frac{1}{4T}
\end{equation}

We show the WDW patch at $t_{0}>t_{c}$ in the Penrose diagram for the MT model in Fig. (\ref{AnisotropicWDW}) (b) and for the DK model in Fig. (\ref{MagneticWDW}) (b). The definition of regions I, II and IV is still given by \eqref{WDWRegionsEarly}, while the one for region III changes to
\begin{equation}
III=\{(t,r)|r\in[r_{m}(t_{0}),r_{h}],\,t\in[t_{0}+r_{\star}(r),-t_{0}-r_{\star}(r)]\}.
\label{WDWRegionsLatter}
\end{equation}

\subsection{MT model}

\begin{figure}[ht!]
 \centering
 \includegraphics[width=0.4\textwidth]{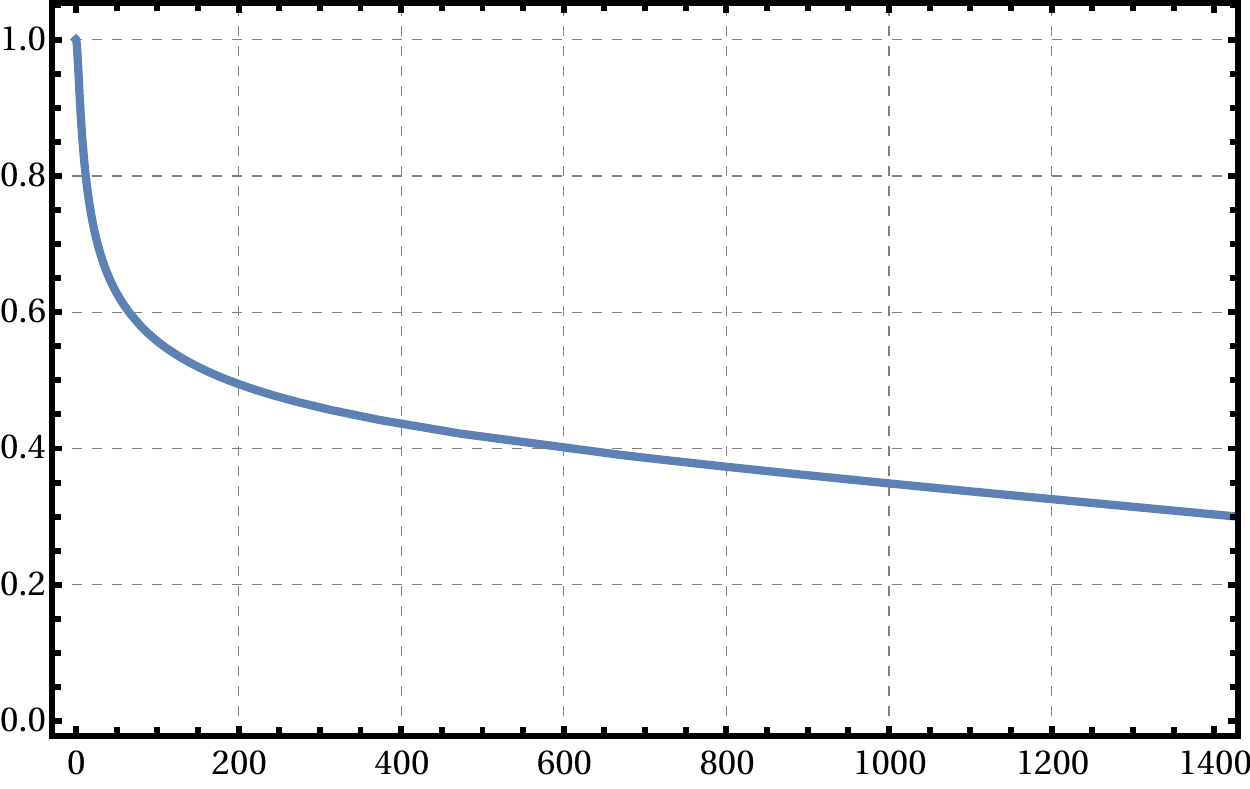}
\put(-0,-10){\Large $\frac{a}{T}$}
 \put(-220,135){\Large $\frac{t_{c}(a)}{t_{c}(0)}$}
\caption{\small Critical time $t_{c}(a)/t_{c}(0)$ as a function of $a/T$ for the MT model.}
\label{Anisotropictc}
\end{figure}

The behaviour of the critical time as a function of the anisotropic parameter $a/T$ is shown in Fig. (\ref{Anisotropictc}). From this it can be seen that $t_{c}(a)/t_{c}(0)$ decreases as $a/T$ increases or, in other words, that the anisotropy has the effect of making the WDW patch withdraw from the past singularity at earlier times compared to the $a=0$ case. This result coincides with what was found in \cite{HosseiniMansoori:2018gdu} for $a/T\ll 1$. 

\begin{figure*}
\begin{center}
\begin{tabular}{cc}
\includegraphics[width=0.45\textwidth]{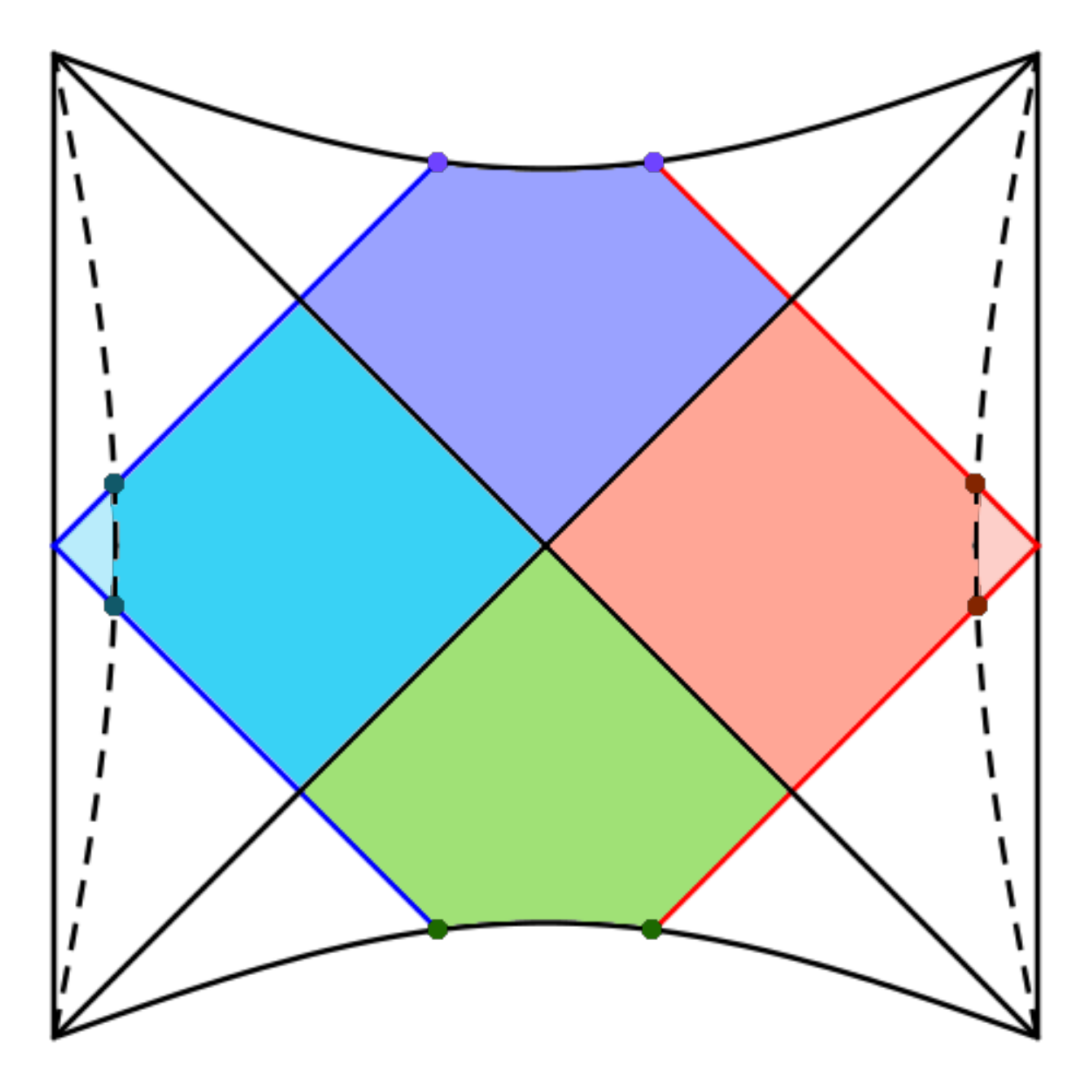} 
\qquad\qquad & 
\includegraphics[width=0.45\textwidth]{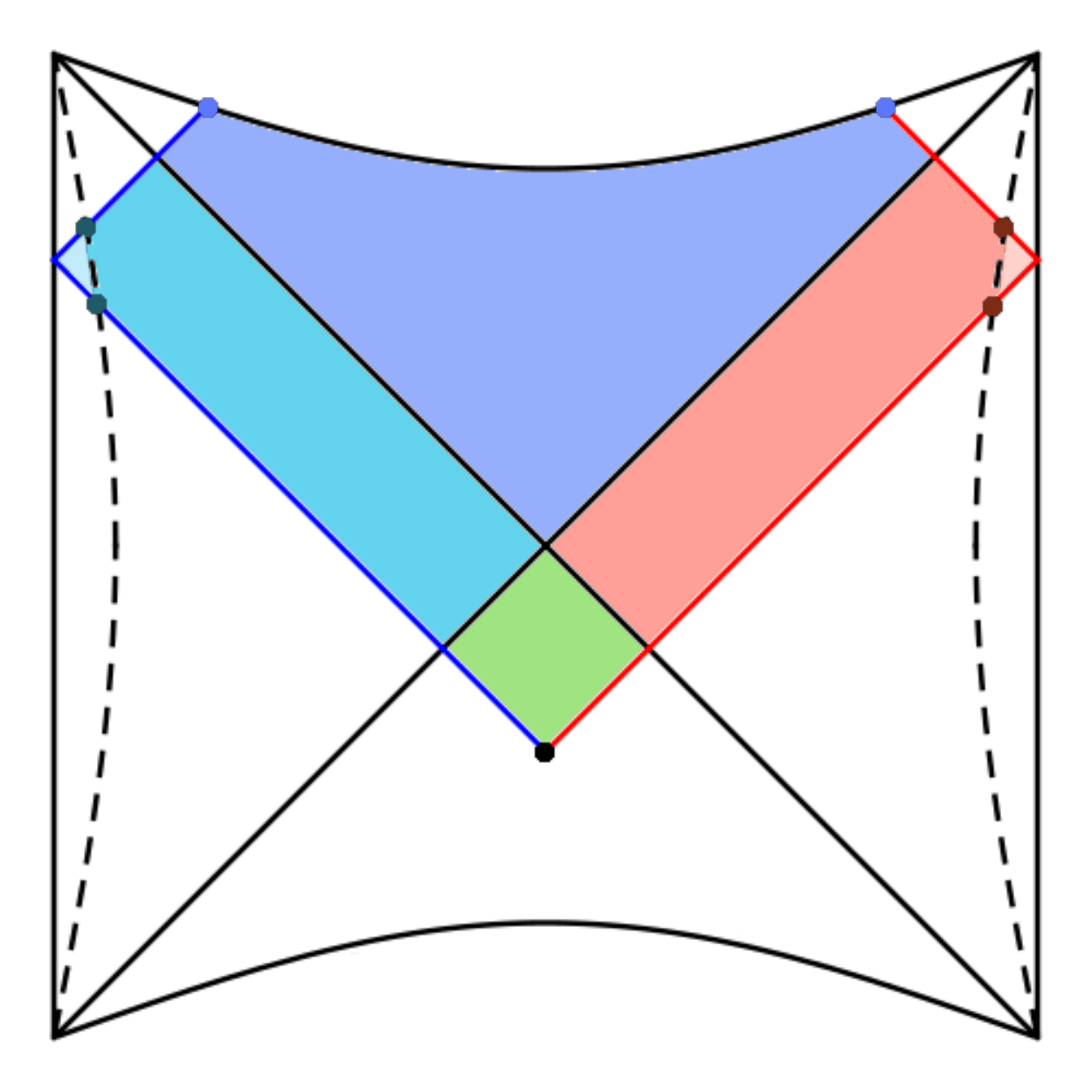}
\qquad
\put(-5,30){\rotatebox{90}{$r=\infty$}}
\put(-225,30){\rotatebox{90}{$r=\infty$}}
\put(-135,200){$r=r_{s}$}
\put(-135,10){$r=r_{s}$}
\put(-5,175){$t_{0}$}
\put(-230,175){$t_{0}$}
\put(-120,140){I}
\put(-90,110){II}
\put(-120,90){III}
\put(-150,110){IV}
\put(-260,30){\rotatebox{90}{$r=\infty$}}
\put(-485,30){\rotatebox{90}{$r=\infty$}}
\put(-380,200){$r=r_{s}$}
\put(-380,10){$r=r_{s}$}
\put(-260,115){$t_{0}$}
\put(-490,115){$t_{0}$}
\put(-370,140){I}
\put(-350,110){II}
\put(-375,90){III}
\put(-400,110){IV}
 \\
(a) & (b)\\
\end{tabular}
\end{center}
\caption{\small Penrose diagram and WDW patch for the MT model. (a) Corresponds to the configuration with $t_{0}/t_{c}=0$, showing that the WDW patch reaches both the future and past singularities. (b) Corresponds to a late time configuration with $t_{0}/t_{c}=1.86$, and it can be seen that the WDW patch no longer reaches the past singularity. For both cases the red and blue lines represent the right and left light sheets respectively, the points denote the joints, while the dashed curve corresponds to the boundary regulator at $r=r_{max}$. The four regions of the WDW patch are displayed: I (dark blue) is the future interior, II (red) is the right exterior, III (green) is the past interior and IV (light blue) is the left exterior. In both figures the anisotropy is $a/T=242$.}
\label{AnisotropicWDW}
\end{figure*}
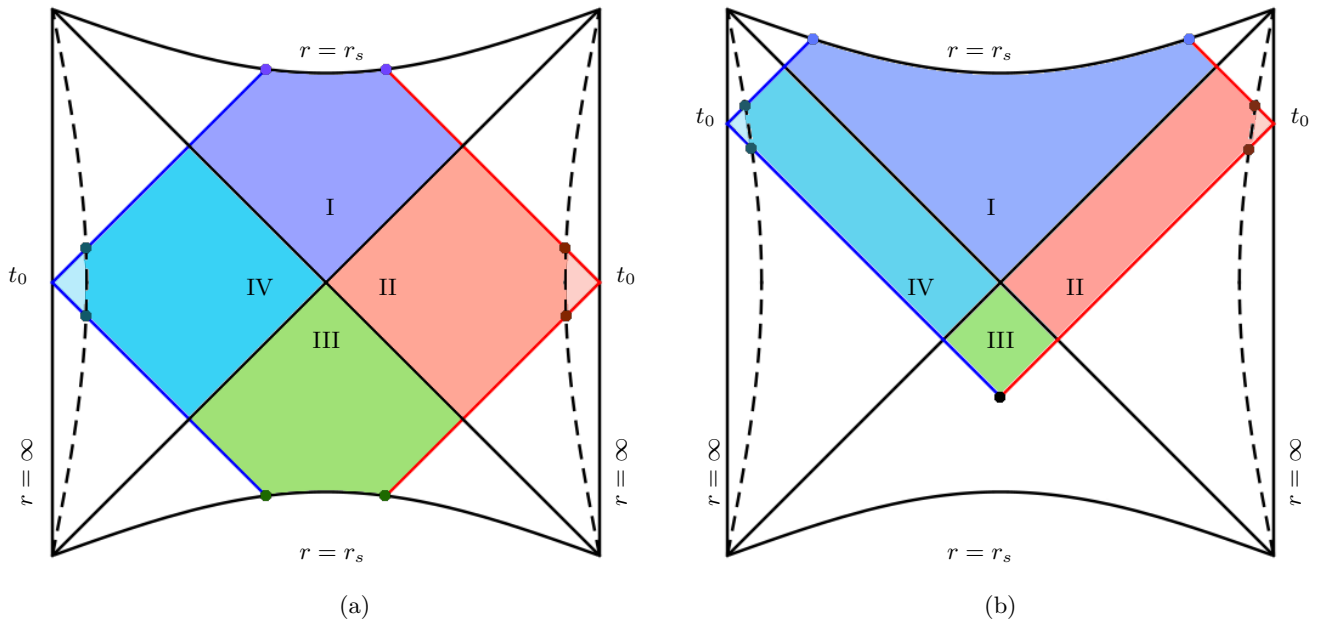

In Fig. (\ref{AnisotropicWDW}) we present the Penrose diagram and the WDW patch for the MT model at $a/T=242$. Fig. (\ref{AnisotropicWDW}) (a) corresponds to $t_{0}/t_{c}=0$, which is an early time configuration because it satisfies $t_{0}<t_{c}$, while Fig. (\ref{AnisotropicWDW}) (b) corresponds to $t_{0}/t_{c}=1.86$, which is a late time configuration with $t_{0}>t_{c}$. In both cases the red and blue lines represent the right and left light sheets respectively, the points denote the joints, while the dashed curves corresponds to the boundary regulator at $r=r_{max}$. We also show explicitly the four regions defined in \eqref{WDWRegionsEarly} and \eqref{WDWRegionsLatter}: I (dark blue) is the future interior, II (red) is the right exterior, III (green) is the past interior and IV (light blue) is the left exterior. It can be seen that in Fig. (\ref{AnisotropicWDW}) (a) the WDW patch reaches the past singularity, and thus there is a total of eight joints to be consider in the evaluation of the holographic complexity. On the other hand, Fig. (\ref{AnisotropicWDW}) (b) shows that for $t_{0}>t_{c}$ the WDW patch no longer reaches the past singularity, and thus there is now a total of seven joints.
\subsection{DK model}

\begin{figure}[ht!]
 \centering
 \includegraphics[width=0.4\textwidth]{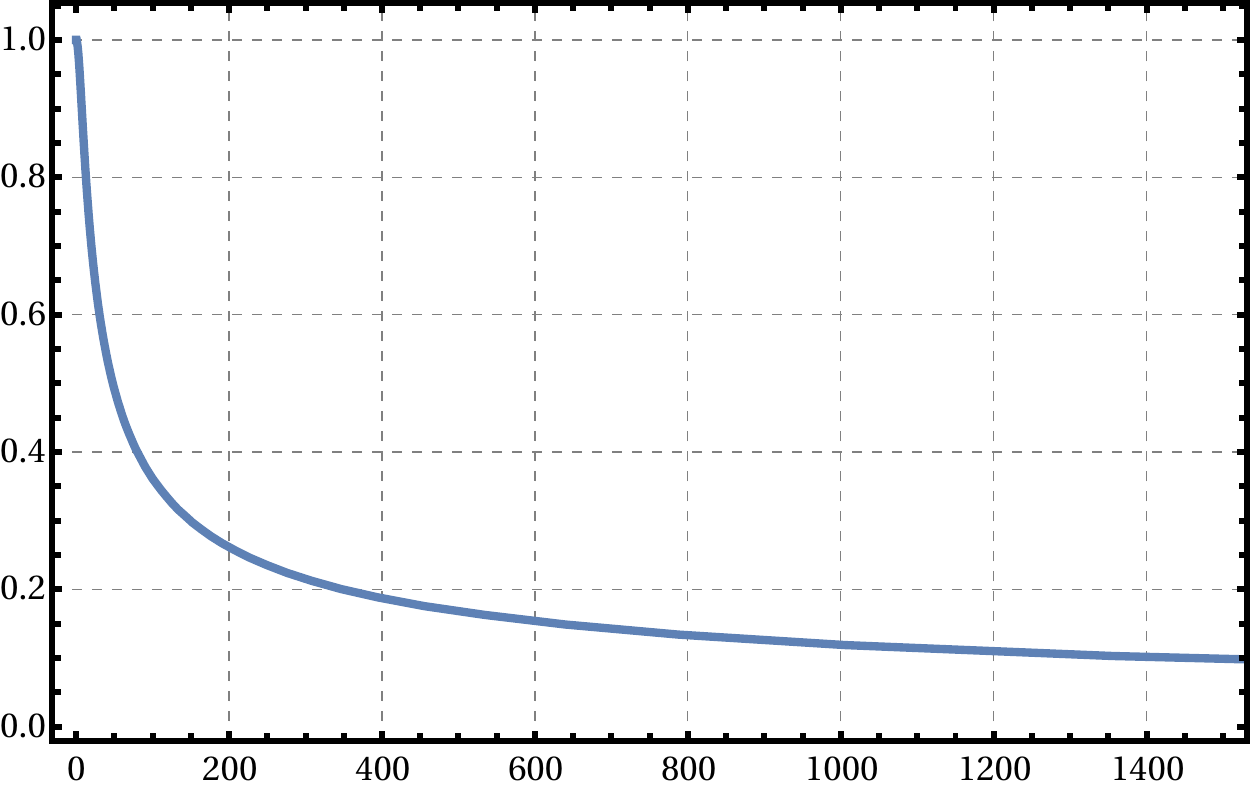}
\put(-0,-10){\Large $\frac{b}{T^{2}}$}
 \put(-230,130){\Large $\frac{t_{c}(b)}{t_{c}(0)}$}
\caption{\small Critical time $t_{c}(b)/t_{c}(0)$ as a function of $b/T^{2}$ for the DK model.}
\label{Magnetictc}
\end{figure}

We show in Fig. (\ref{Magnetictc}) how the critical time depends on the magnetic field intensity $b/T^{2}$ for the DK model. From this it can be seen that $t_{c}(b)/t_{c}(0)$ decreases as $b/T^{2}$ increases, that is, similarly to the anisotropy in the MT model, the magnetic field has the effect of making the WDW patch withdraw from the past singularity at earlier times compared to the $b=0$ case.

\begin{figure*}
\begin{center}
\begin{tabular}{cc}
\includegraphics[width=0.45\textwidth]{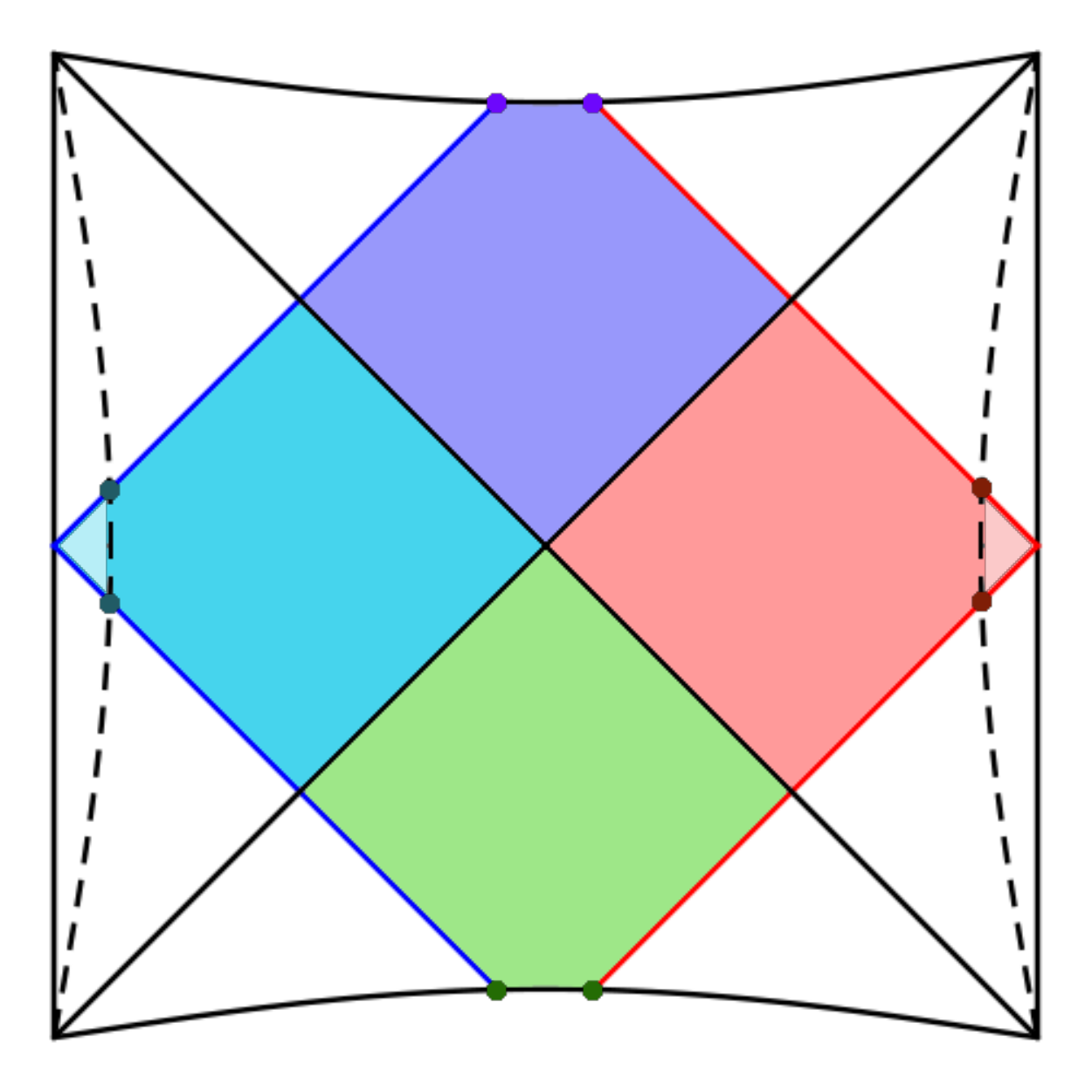} 
\qquad\qquad & 
\includegraphics[width=0.45\textwidth]{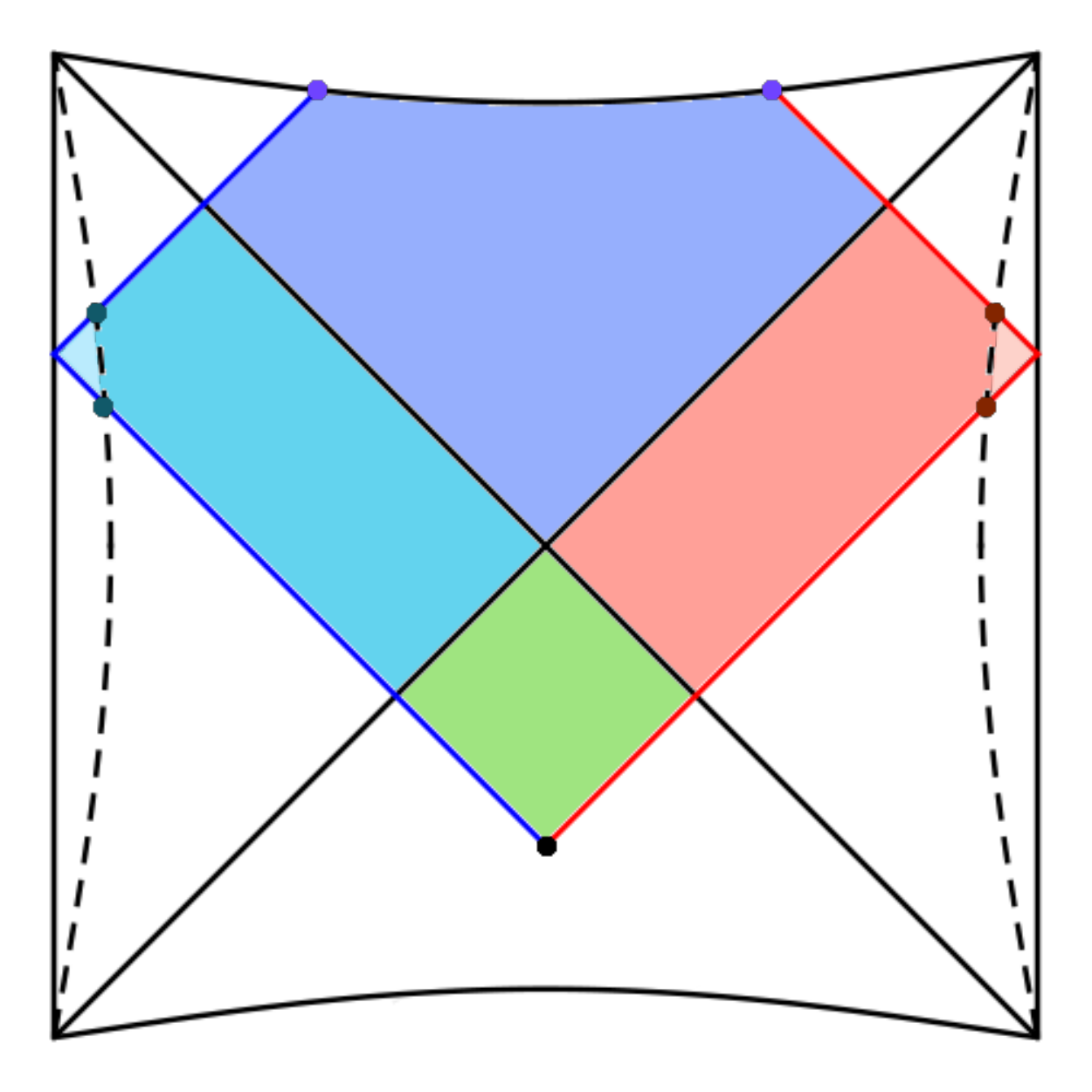}
\qquad
\put(-5,30){\rotatebox{90}{$r=\infty$}}
\put(-225,30){\rotatebox{90}{$r=\infty$}}
\put(-135,215){$r=r_{s}$}
\put(-135,10){$r=r_{s}$}
\put(-5,155){$t_{0}$}
\put(-230,155){$t_{0}$}
\put(-120,140){I}
\put(-90,110){II}
\put(-120,90){III}
\put(-150,110){IV}
\put(-260,30){\rotatebox{90}{$r=\infty$}}
\put(-485,30){\rotatebox{90}{$r=\infty$}}
\put(-380,215){$r=r_{s}$}
\put(-380,10){$r=r_{s}$}
\put(-260,115){$t_{0}$}
\put(-490,115){$t_{0}$}
\put(-370,140){I}
\put(-350,110){II}
\put(-375,90){III}
\put(-400,110){IV}
 \\
(a) & (b)\\
\end{tabular}
\end{center}
\caption{\small Penrose diagram and WDW patch for the DK model. (a) Corresponds to an early time configuration with $t_{0}/t_{c}=0$, as the WDW patch reaches both the future and past singularities. (b) Corresponds to a late time configuration with $t_{0}/t_{c}=3.22$, as the WDW patch no longer reaches the past singularity. For both cases the red and blue lines represent the right and left light sheets respectively, the points denote the joints, while the dashed curve corresponds to the boundary regulator at $r=r_{max}$. The four regions of the WDW patch are displayed: I (dark blue) is the future interior, II (red) is the right exterior, III (green) is the past interior and IV (light blue) is the left exterior. In both figures the magnetic field intensity is $b/T^{2}=152.20$, which means that the singularity is located at $r_{s}/r_{h}=-0.06$.}
\label{MagneticWDW}
\end{figure*}

In Fig. (\ref{MagneticWDW}) we show the Penrose diagram and the WDW patch for the DK model at $b/T^{2}=152.20$. Fig. (\ref{MagneticWDW}) (a) corresponds to an early time $t_{0}<t_{c}$ given by $t_{0}/t_{c}=0$, while Fig. (\ref{MagneticWDW}) (b) corresponds to $t/t_{c}=3.22$, which is a late time configuration with $t_{0}>t_{c}$. Note that, as mentioned in Sec. \ref{MagneticBackground}, the position of the singularity in the $r$-coordinate changes with the magnetic field. For $b/T^{2}=152.20$, the singularity is located at $r_{s}/r_{h}=-0.06$. As before, the red and blue lines represent the right and left light sheets respectively, the points denote the joints, while the dashed curves corresponds to the boundary regulator at $r=r_{max}$. The Penrose diagram conveniently displays the four regions defined in \eqref{WDWRegionsEarly} and \eqref{WDWRegionsLatter}: I (dark blue) is the future interior, II (red) is the right exterior, III (green) is the past interior and IV (light blue) is the left exterior. Just as for the MT model, it can be seen from Fig. (\ref{MagneticWDW}) (a) that the WDW patch reaches the past singularity, and thus there is a total of eight joints to consider in the evaluation of the holographic complexity. On the other hand, Fig. (\ref{MagneticWDW}) (b) shows that for $t_{0}>t_{c}$ the WDW patch no longer reaches the past singularity, and thus there is now a total of seven joints.
\section{Holographic complexity}
\label{Complexity}

In this section we compute the holographic complexity using the CA prescription. We follow closely the analysis presented in \cite{HosseiniMansoori:2018gdu}, but offering additional details regarding the particularities of our numerical solutions, such as the conformal anomaly. As previously explained, the evaluation of the action over the WdW patch requires the addition of extra boundary terms to make sure that the variational principle is well-posed. Thus the full action has three types of contributions: one related to the bulk, other related to boundary surfaces, and the last one for the joints between these surfaces, i.e., 
\begin{equation}
S_{WDW}=S_{bulk}+S_{surface}+S_{joint},
\end{equation}
where the schematic form of each term is
\begin{eqnarray}
&& S_{bulk}=\frac{1}{16\pi G_{5}}\int_{\mathcal{M}}d^{5}x\sqrt{-g}\mathcal{L},
\cr
&& S_{surface}=\frac{1}{16\pi G_{5}}\int_{\mathcal{B}} d^{4}x\sqrt{|\gamma|}2K+\frac{1}{16\pi G_{5}}\int_{\mathcal{B}'}d\lambda d^{3}\theta\sqrt{h}2\kappa,
\cr
&& S_{joint}= \frac{1}{16\pi G_{5}}\int_{\Sigma'}d^{3}x\sqrt{\sigma}2 a.
\label{IntegralDefinitions}
\end{eqnarray}

The first expression in \eqref{IntegralDefinitions} is the bulk action of the theory, where $\mathcal{M}$ refers to the interior of the WdW patch. The first integral on the right hand side of the second line of \eqref{IntegralDefinitions} is the YGH surface term for the spacelike and timelike segments $\mathcal{B}$ of the boundary of the WDW patch, which is written using the induced metric $\gamma$ and the extrinsic curvature $K$. The second integral on the same expression is the surface term for the null segments $\mathcal{B}'$ of the boundary of the WDW patch, defined by the induced metric $h$ and the function $\kappa$, which measures the failure of the null generators to be affinely parametrized. Lastly, $S_{joint}$ denote the joint terms, where $\sigma$ is the determinant of the induced metric on this codimension-2 surfaces. The integrand of this expression takes different forms depending on which kind of intersection is considered. For the three possible cases we explicitly have
\begin{equation}
a=\left\{
\begin{array}{lll}
		 \epsilon\log|\boldsymbol{t}\cdot\boldsymbol{k}|, & \text{for space-null joint with} \\ & \epsilon=-\text{sign}(\boldsymbol{t}\cdot\boldsymbol{k})\,\text{sign}(\hat{s}\cdot\boldsymbol{k}),\\
         \epsilon\log|\boldsymbol{s}\cdot\boldsymbol{k}|, &  \text{for time-null joint with} \\ & \epsilon=-\text{sign}(\boldsymbol{s}\cdot\boldsymbol{k})\,\text{sign}(\hat{t}\cdot\boldsymbol{k}),\\
         \epsilon\log\left|\frac{\boldsymbol{k_{1}}\cdot\boldsymbol{k_{2}}}{2}\right|, & \text{for null-null joint with} \\ & \epsilon=-\text{sign}(\boldsymbol{k_{1}}\cdot\boldsymbol{k_{2}})\,\text{sign}(\hat{k}_{1}\cdot\boldsymbol{k_{2}}),
\end{array}
\right.
\label{JointIntegrand}
\end{equation}
where $\boldsymbol{s}$, $\boldsymbol{t}$ and $\boldsymbol{k}$ are the normalized outward unit normal one-forms to the spacelike, timelike or null surface under consideration respectively, while $\boldsymbol{k_{1}}$ and $\boldsymbol{k_{2}}$ are two normal outward one-forms to different but intersecting null surfaces. The sign $\epsilon$ in each case is determined by the auxiliary vectors $\hat{s}$, $\hat{t}$, $\hat{k_{1}}$, normal to the spacelike, timelike and null surfaces respectively. We will explicitly write this vectors and one-forms below, following the procedure and conventions outlined in \cite{Carmi:2016wjl}.

As first explained in \cite{Lehner:2016vdi}, there are certain ambiguities associated with the null surface contributions. In particular, the action depends on the parametrization chosen for the null generators as, for example, an affine parametrization sets $\kappa=0$ and eliminates the null surfaces contributions from the action altogether. In order to remove this ambiguity, it is necessary to add the counterterm
\begin{equation}
S_{null}=\frac{1}{16\pi G_{5}}\int d\lambda d^{3}\theta \sqrt{h} 2\Theta\log(l_{null}\Theta),
\label{NullCounter}
\end{equation}  
where $\Theta$ is the expansion of the null generators, defined as 
\begin{equation}
\Theta=\partial_{\lambda}\log(\sqrt{h}).
\label{Theta}
\end{equation}
While \eqref{NullCounter} is by itself dependent on the parametrization of the null generators, the full action turns out invariant when adding it. Thus we will follow usual conventions and chose an affine $\lambda$. Also note that $S_{null}$ introduces an arbitrary lenght scale $l_{null}$, which is related to the freedom of choosing a reference state in the dual theory \cite{Akhavan:2019zax}. We will leave it arbitrary and show that it does not modify the late time behavior of the rate of change of the complexity for the type of geometries that we consider. 

The inclusion of this term does not affect certain aspects of the complexity, such as the complexity of formation \cite{Chapman:2016hwi} or the late time behavior of the complexity in the case of the eternal black hole \cite{Carmi:2017jqz}. However, it can modify other aspects\footnote{This term is crucial when the spacetime under consideration is not stationary, such as Vaidya spacetime \cite{Chapman:2018dem,Chapman:2018lsv}} like the structure of the UV divergences \cite{Reynolds:2016rvl,Akhavan:2019zax}. Here we will consider $S_{null}$ to determine if it is affected by the presence of the conformal anomaly.

\subsection{Before the critical time: $\boldsymbol{0\leq t_{0}\leq t_{c}}$}

We first compute the bulk contribution. In order to do this, we need to evaluate the action on each of the four regions that constitute the WDW patch
\begin{equation}
S_{bulk}=S_{I}+S_{II}+S_{III}+S_{IV}.
\label{Sbulk1}
\end{equation}
Given that the metric and the Lagrangian only depend on $r$, when evaluating $S_{bulk}$ we can integrate along the $(x,y,z)$ directions and obtain an overall factor of $V_{x}$. The integration over time is also easily performed using the definition of each region given in \eqref{WDWRegionsEarly}. Thus we have
\begin{eqnarray}
&& S_{I}=2\mathcal{N}\int_{r_{min}}^{r_{h}}dr\sqrt{-g}\mathcal{L}(r)(t_{0}-r_{\star}(r)),
\cr
&& S_{II}=S_{IV}=-2\mathcal{N}\int_{r_{h}}^{r_{max}}dr\sqrt{-g}\mathcal{L}(r)r_{\star}(r),
\cr
&& S_{III}=2\mathcal{N}\int_{r_{min}}^{r_{h}}dr\sqrt{-g}\mathcal{L}(r)(-t_{0}-r_{\star}(r)),
\end{eqnarray}
and by substituting this into \eqref{Sbulk1} we obtain
\begin{equation}
S_{bulk}(0\leq t_{0}\leq t_{c})=-4\mathcal{N}\int_{r_{min}}^{r_{max}}dr\sqrt{-g}\mathcal{L}(r)r_{\star}(r).
\label{SbulkEarly}
\end{equation}
At this stage we note two important things. The first one is that the $t_{0}$ dependence has been completely eliminated from \eqref{SbulkEarly}. This is because the volume of the WDW patch remains constant for any $0\leq t_{0}\leq t_{c}$ and any member of the family of solutions of either the MT and DK model is static. The second one is that this integral is divergent as the limit $r_{max}\rightarrow\infty$ is taken.

Next we turn to the surface integrals. As previously explained, by choosing an affine parameter we eliminate the contribution coming from the second term in the second line of \eqref{IntegralDefinitions}. Hence we are left with the integrals at the two spacelike and the two timelike surfaces at $r=r_{max}$ and $r=r_{min}$ respectively. Schematically we have
\begin{equation}
S_{surface}=S_{future}+S_{right}+S_{past}+S_{left},
\label{Ssurf1}
\end{equation}
where $S_{future}$ denotes the integral evaluated at the regulator near the future singularity, $S_{past}$ is the analogous term for the past singularity, and $S_{right}$ and $S_{left}$ correspond to the integrals at the regulators near the right and left boundaries respectively. In all cases the integral to consider is the YGH term, given by the induced metric in the surface $\gamma_{ij}$ and the extrinsic curvature
\begin{equation}
K_{ij}=\frac{\partial x^{\mu}}{\partial y^{i}}\frac{\partial x^{\nu}}{\partial y^{j}}\nabla_{\mu}n_{\nu}, \qquad K=\gamma^{ij}K_{ij},
\end{equation}
where $y^{i}$ are the coordinates on the surface defined by $x^{\mu}=x^{\mu}(y^{i})$, and $n_{\mu}$ are the components of the corresponding unit outward normal one-form. The latter are explicitly given by
\begin{eqnarray}
&& \boldsymbol{s}=\sqrt{|g_{rr}|}\,dr, \qquad \text{for} \qquad r=r_{max},
\cr
&& \boldsymbol{t}=-\sqrt{|g_{rr}|}\,dr, \qquad \text{for} \qquad r=r_{min}.
\label{spacetime1form}
\end{eqnarray}

Applying the previous expressions for both the MT and DK models, it can be shown that the integrand of the YGH term is a function of $r$ only
\begin{equation}
2\sqrt{|\gamma|}K=\pm \mathcal{G}(r),
\end{equation}
where the plus sign is used for the $r=r_{max}$ surfaces and the minus sign is used for the near singularity regulators at $r=r_{min}$. Using this fact we can easily evaluate the surface integrals, which gives
\begin{eqnarray}
&& S_{future}=-2\mathcal{N}\mathcal{G}(r_{min})(t_{0}-r_{\star}(r_{min})),
\cr
&& S_{right}=S_{left}=-2\mathcal{N}\mathcal{G}(r_{max})r_{\star}(r_{max}),
\cr
&& S_{past}=-2\mathcal{N}\mathcal{G}(r_{min})(-t_{0}-r_{\star}(r_{min})),
\end{eqnarray}
and thus after substitution in \eqref{Ssurf1} the result is
\begin{equation}
S_{surface}(0\leq t_{0}\leq t_{c})=-4\mathcal{N}\mathcal{G}(r)r_{\star}(r)\bigg|^{r_{max}}_{r_{min}}.
\label{SsurfEarly}
\end{equation}
Note that once again, this contribution is independent of $t_{0}$. This is because the $r_{max}$ surfaces only suffer a time translation, while the area that the future $r_{min}$ surface gains is exactly the area that the past $r_{min}$ one looses. In other words, the $t_{0}$ dependence of $S_{future}$ exactly cancels the one coming from $S_{past}$. Also note that $S_{surface}$ diverges as the limit $r_{max}\rightarrow\infty$ is taken.

Next we have the contribution coming from the joint terms. As can be seen from Fig. (\ref{AnisotropicWDW}) (a) and Fig. (\ref{MagneticWDW}) (a), there are eight joints to consider for $0\leq t_{0}\leq t_{c}$, and all of them are of the space-null or the time-null type. In order to evaluate $S_{joint}$ we need to define the auxiliary vectors appearing in \eqref{JointIntegrand}. As explained in \cite{Carmi:2016wjl}, this vectors need to live in the tangent space of the corresponding surface, be ortogonal to the joint, and point outward to the WDW patch. Note that this vectors do not need to be normalized, as can be seen from the way they enter in \eqref{JointIntegrand}. For the space-null joints we have
\begin{equation}
\hat{s}_{\pm}=\pm\partial_{t},
\label{spacevector}
\end{equation}
where the plus sign is used for the joints at the future singularity regulator, and the minus sign for the ones at the past singularity regulator. On the other hand, for the time-null joints we use 
\begin{equation}
\hat{t}_{\pm}=\pm\partial_{t},
\label{timevector}
\end{equation}
where the plus sign is used for the joints at the future light sheets, while the minus sign is used for the joints at the past light sheets. We also need the outward normal one-forms to the null hypersurfaces, which are specified by an expression of the form $\Phi(x^{\mu})=0$ (explicitly given in \eqref{LightSheets}). Thus the normal one-forms are given by $\alpha\,d\Phi$, where $\alpha>0$ is an arbitrary but positive normalization constant. While this is another ambiguity that comes from the fact that the WDW patch contains null segments, as we will see below this arbitrary constant affects the value of the complexity of the state, but the late time behavior of the rate of change of the complexity is unaffected by it (see \cite{Lehner:2016vdi,Carmi:2016wjl,Carmi:2017jqz} for a detailed discussion). For each light sheet we have
\begin{equation}
\boldsymbol{k_{1}}=\alpha\left(\text{sign}(g_{rr})\sqrt{\left|\frac{g_{rr}}{g_{tt}}\right|}\,dr+dt\right),
\end{equation}
for the right future and left past light sheet and
\begin{equation}
\boldsymbol{k_{2}}=\alpha\left(\text{sign}(g_{rr})\sqrt{\left|\frac{g_{rr}}{g_{tt}}\right|}\,dr-dt\right),
\label{Null1forms}
\end{equation}
for the right past and left future light sheet.

After substitution of \eqref{spacetime1form}, \eqref{spacevector}, \eqref{timevector} and \eqref{Null1forms} into \eqref{JointIntegrand}, for the joints in regions I and III we have
\begin{equation}
S_{joint}=\left.2\mathcal{N}\sqrt{\sigma}\log\left(\frac{\alpha}{\sqrt{|g_{tt}|}}\right)\right|_{r=r_{min}},
\end{equation}
while for the joints in regions II and IV
\begin{equation}
S_{joint}=-\left.2\mathcal{N}\sqrt{\sigma}\log\left(\frac{\alpha}{\sqrt{|g_{tt}|}}\right)\right|_{r=r_{max}},
\end{equation}
hence the full joint contribution is
\begin{equation}
S_{joint}(0\leq t_{0}\leq t_{c})=-8\mathcal{N}\sqrt{\sigma}\log\left(\frac{\alpha}{\sqrt{|g_{tt}|}}\right)\bigg|^{r_{max}}_{r_{min}}.
\label{SjointEarly}
\end{equation}
Note that also $S_{joint}$ is independent of $t_{0}$.

Finally we have the contribution coming from the null counterterm \eqref{NullCounter}. The generators for each hypersurface are given by \eqref{LightSheets}, hence we parametrize the null surfaces with $\theta^{i}=(x,y,z)$ and the induced metric is given by
\begin{equation}
h_{AB}=g_{\alpha\beta}\frac{dx^{\alpha}}{d\theta^{A}}\frac{dx^{\beta}}{d\theta^{B}}.\label{hABnull}
\end{equation}
From \eqref{hABnull} we see that for metrics with the structure \eqref{PenroseMetric} that we are considering, $h_{AB}$ is equal to the induced metric on the joints $\sigma_{AB}$. To make the above consistent with our previous calculations, we need to chose an affine parameter for the null generators. A direct substitution of the normal vectors as written in \eqref{Null1forms} in the geodesic equationd yields
\begin{equation}
k^{\alpha}\nabla_{\alpha}k_{\mu}=0,
\end{equation}
for any constant $\alpha>0$. Hence, using the fact that
\begin{equation}
k^{\mu}=\frac{d x^{\mu}}{d\lambda},
\end{equation}
the affine parameter $\lambda$ is given by the solution to the equations
\begin{equation}
\frac{dr}{d\lambda}=\frac{\alpha}{\sqrt{|g_{rr}g_{tt}|}}, \qquad \frac{dt}{d\lambda}=\pm\frac{\alpha}{g_{tt}},
\label{AffineParameter}
\end{equation}
in which derivation we have used \eqref{rmderivative}.

With the previous choices, the integration over $\theta^{A}$ in \eqref{NullCounter} gives a factor of $V_{x}$, while we can employ \eqref{AffineParameter} to perform a change of variables such that the integration over $\lambda$ is converted to be over $r$, resulting in
\begin{equation}
S_{null}=\frac{2\mathcal{N}}{\alpha}\int dr\sqrt{|g|}\Theta\log(l_{null}\Theta).
\label{null1}
\end{equation}
We can also use \eqref{AffineParameter} to evaluate $\Theta$ applying the chain rule, which gives
\begin{equation}
\Theta=\frac{\alpha}{\sqrt{|g|}}\partial_{r}\sqrt{\sigma}.
\end{equation}
Substitution of this last expression further simplifies \eqref{null1} into
\begin{equation}
S_{null}=2\mathcal{N}\int dr \log\left(\frac{\alpha\,l_{null}}{\sqrt{|g|}}\partial_{r}\sqrt{\sigma}\right)\partial_{r}\sqrt{\sigma}.
\label{NullFinal}
\end{equation}

The crucial point extracted from the previous calculations is that the integrand is independent of $t_{0}$. For any $0\leq t_{0}\leq t_{c}$ the integration over the four light sheets runs from $r_{min}$ to $r_{max}$
\begin{equation}
S_{null}(0\leq t_{0}\leq t_{c})=8\mathcal{N}\int_{r_{min}}^{r_{max}}dr \log\left(\frac{\alpha\,l_{null}}{\sqrt{|g|}}\partial_{r}\sqrt{\sigma}\right)\partial_{r}\sqrt{\sigma},
\label{NullEarly}
\end{equation}
hence $S_{null}$ does not depend on $t_{0}$. 

We thus conclude that $S_{WDW}$, and as a consequence the complexity of the state, is constant for $0\leq t_{0}\leq t_{c}$.

\subsection{After the critical time: $\boldsymbol{t_{0}>t_{c}}$}

As before, we start by evaluating the bulk contribution in \eqref{IntegralDefinitions} using the definition of each region given in \eqref{WDWRegionsEarly} and \eqref{WDWRegionsLatter}. We have
\begin{eqnarray}
&& S_{I}=2\mathcal{N}\int_{r_{min}}^{r_{h}}dr\sqrt{-g}\mathcal{L}(r)(t_{0}-r_{\star}(r)),
\cr
&& S_{II}=S_{IV}=-2\mathcal{N}\int_{r_{h}}^{r_{max}}dr\sqrt{-g}\mathcal{L}(r)r_{\star}(r),
\cr
&& S_{III}=2\mathcal{N}\int_{r_{m}}^{r_{h}}dr\sqrt{-g}\mathcal{L}(r)(-t_{0}-r_{\star}(r)).
\end{eqnarray}
Note that the $t_{0}$ dependence in $S_{III}$ is no longer cancelled by $S_{I}$. To see explicitly how this happens let us rewrite $S_{III}$ as
\begin{equation}
\begin{split}
S_{III}=&-2\mathcal{N}\int_{r_{min}}^{r_{m}}dr\sqrt{-g}\mathcal{L}(r)(-t_{0}-r_{\star}(r))\\&+2\mathcal{N}\int_{r_{min}}^{r_{h}}dr\sqrt{-g}\mathcal{L}(r)(-t_{0}-r_{\star}(r)),
\end{split}
\end{equation}
where we used the fact that $r_{min}<r_{m}$. Thus, using \eqref{SbulkEarly}, the bulk action evaluated at the WDW patch for latter times can be written as
\begin{equation}
\begin{split}
S_{bulk}(t_{0}>t_{c})= & S_{bulk}(0\leq t_{0}\leq t_{c}) \\ &+2\mathcal{N}\int_{r_{mim}}^{r_{m}}dr\sqrt{-g}\mathcal{L}(r)(t_{0}+r_{\star}(r)).
\end{split}
\label{SbulkLatter}
\end{equation}
Note that $S_{bulk}(t_{0}>t_{c})$ diverges as the regulator $r_{max}$ is taken to infinity, but this behavior is contained in the $t_{0}$-independent term $S_{bulk}(t_{0}<t_{c})$.

Next we turn to the surface integrals. Once again we set the contribution of the null hypersurfaces to zero by choosing an affine parameter. As the WDW patch no longer reaches the past singularity, we are left with
\begin{equation}
S_{surface}=S_{future}+S_{left}+S_{right}.
\end{equation}
Using the previous expressions for the extrinsic curvature and induced metric on the hypersurfaces, and the definition of the regions, we can compute the corresponding integrals. For each one we have
\begin{eqnarray}
&& S_{future}=-2\mathcal{N}\mathcal{G}(r_{min})(t_{0}-r_{\star}(r_{min})),
\cr
&& S_{right}=S_{left}=-2\mathcal{N}\mathcal{G}(r_{max})r_{\star}(r_{max}),
\end{eqnarray}
and thus the total surface action is
\begin{equation}
\begin{split}
S_{surface}(t_{0}>t_{c})= & -4\mathcal{N}\mathcal{G}(r_{max})r_{\star}(r_{max}) \\ &-2\mathcal{N}\mathcal{G}(r_{min})(t_{0}-r_{\star}(r_{min})).
\end{split}
\end{equation}
Note that the $t_{0}$ dependence coming from the integration at the future singularity regulator is no longer canceled. Explicitly we have that
\begin{equation}
\begin{split}
S_{surface}(t_{0}>t_{c})= & S_{surface}(0\leq t_{0}\leq t_{c}) \\& -2\mathcal{N}\mathcal{G}(r_{min})(t_{0}+r_{\star}(r_{min})),
\end{split}
\label{SsurfLatter}
\end{equation}
where the $r_{max}\rightarrow\infty$ divergence is contained in $S_{surface}(0\leq t_{0}\leq t_{c})$, which is independent of $t_{0}$. 

Next we have the contribution coming from the joints. For latter times we have only 7 joints to consider, from which 6 are of the space-null and time-null type, as can be seen in Fig. (\ref{AnisotropicWDW}) (b) and Fig. (\ref{MagneticWDW}) (b). We can evaluate those using the 1-forms and auxiliary vectors defined in the previous subsection. For the joints in region I we have
\begin{equation}
S_{joint}=\left.2\mathcal{N}\sqrt{\sigma}\log\left(\frac{\alpha}{\sqrt{|g_{tt}|}}\right)\right|_{r=r_{min}},
\end{equation}
while for the joints in regions II and IV
\begin{equation}
S_{joint}=-\left.2\mathcal{N}\sqrt{\sigma}\log\left(\frac{\alpha}{\sqrt{|g_{tt}|}}\right)\right|_{r=r_{max}}.
\end{equation}
In order to evaluate the contribution coming from the null-null joint, we need to define the auxiliary vector appearing in \eqref{JointIntegrand}. Following \cite{Carmi:2016wjl} we have
\begin{equation}
\hat{k_{1}}=\text{sign}(g_{rr})\sqrt{\left|\frac{g_{rr}}{g_{tt}}\right|}\partial_{r}-\partial_{t},
\end{equation}
which lives on the left past light sheet. Substitution of this in \eqref{JointIntegrand} gives
\begin{equation}
S_{null-null}=\left.2\mathcal{N}\sqrt{\sigma}\log\left(\frac{\alpha^{2}}{|g_{tt}|}\right)\right|_{r=r_{m}}.
\end{equation}
After adding all the joint contributions we have
\begin{equation}
\begin{split}
S_{joint}(t_{0}>t_{c})=&-\left.8\mathcal{N}\sqrt{\sigma}\log\left(\frac{\alpha}{\sqrt{|g_{tt}|}}\right)\right|_{r=r_{max}} \\& +\left.4\mathcal{N}\sqrt{\sigma}\log\left(\frac{\alpha}{\sqrt{|g_{tt}|}}\right)\right|_{r=r_{min}}\\&+\left.2\mathcal{N}\sqrt{\sigma}\log\left(\frac{\alpha^{2}}{|g_{tt}|}\right)\right|_{r=r_{m}}.
\end{split}
\end{equation}
Note that the $t_{0}$ dependence is implicit in $r_{m}$. To isolate this time dependence we can rewrite the previous expression as
\begin{equation}
\begin{split}
S_{joint}(t_{0}>t_{c})=&S_{joint}(0\leq t_{0}\leq t_{c}) \\& -\left.4\mathcal{N}\sqrt{\sigma}\log\left(\frac{\alpha}{\sqrt{|g_{tt}|}}\right)\right|_{r=r_{min}}\\&+\left.2\mathcal{N}\sqrt{\sigma}\log\left(\frac{\alpha^{2}}{|g_{tt}|}\right)\right|_{r=r_{m}}.
\end{split}
\label{SjointLatter}
\end{equation}

Finally we have the contribution coming from the null counterterm. For later times the two past light sheets no longer reach the singularity and end at $r_{m}$ instead. Thus we explicitly have
\begin{equation}
\begin{split}
S_{null}(t_{0}>t_{c})=&4\mathcal{N}\int_{r_{min}}^{r_{max}}dr \log\left(\frac{\alpha\,l_{null}}{\sqrt{|g|}}\partial_{r}\sqrt{\sigma}\right)\partial_{r}\sqrt{\sigma} \\& +4\mathcal{N}\int_{r_{m}}^{r_{max}}dr \log\left(\frac{\alpha\,l_{null}}{\sqrt{|g|}}\partial_{r}\sqrt{\sigma}\right)\partial_{r}\sqrt{\sigma}.
\end{split}
\end{equation}
In order to isolate the $t_{0}$ dependence of the last expression we rewrite the second term as an integral from $r_{min}$ to $r_{max}$ minus an integral from $r_{min}$ to $r_{m}$, which gives
\begin{equation}
\begin{split}
S_{null}(t_{0}>t_{c})=&S_{null}(0\leq t_{0}\leq t_{c}) \\& -4\mathcal{N}\int_{r_{min}}^{r_{m}}dr \log\left(\frac{\alpha\,l_{null}}{\sqrt{|g|}}\partial_{r}\sqrt{\sigma}\right)\partial_{r}\sqrt{\sigma},
\end{split}
\label{NullLatter}
\end{equation}
where we have used \eqref{NullEarly}

We can now proceed to compute $dS_{WDW}/dt_{0}$. By taking the derivative of \eqref{SbulkLatter} with respect to $t_{0}$ we obtain
\begin{equation}
\frac{dS_{bulk}}{dt_{0}}=2\mathcal{N}\int_{r_{min}}^{r_{m}}dr\sqrt{-g}\mathcal{L}(r),
\end{equation}
where the contribution coming from the upper integration limit vanishes because of the equation that defines $r_{m}$ \eqref{rm}. On the other hand, the derivative of the surface integral \eqref{SsurfLatter} gives
\begin{equation}
\frac{dS_{surface}}{dt_{0}}=-2\mathcal{N}\mathcal{G}(r_{min}),
\end{equation}
while the joint contribution is
\begin{equation}
\begin{split}
\frac{dS_{joint}}{dt_{0}}=&\left.2\mathcal{N}\sqrt{-\frac{g_{tt}}{g_{rr}}}\left(\sqrt{\sigma}\frac{g_{tt}'}{g_{tt}}+\frac{\sigma'}{2\sqrt{\sigma}}\log\left(\frac{\alpha^{2}}{g_{tt}}\right)\right)\right|_{r=r_{m}},
\end{split}
\end{equation}
where the prime denotes the derivative with respect to $r$ and we have used the expression for the derivative of $r_{m}$ with respect to $t_{0}$ \eqref{rmderivative}. We also used the fact that $g_{tt}(r_{m})>0$ and $g_{rr}(r_{m})<0$. Finally, the $t_{0}$ derivative of the null counterterm \eqref{NullLatter} is
\begin{equation}
\frac{dS_{null}}{dt_{0}}=-4\mathcal{N}\left.\sqrt{-\frac{g_{tt}}{g_{rr}}}\frac{\sigma'}{2\sqrt{\sigma}}\log\left(\frac{\alpha\,l_{null}}{\sqrt{|g|}}\frac{\sigma'}{2\sqrt{\sigma}}\right)\right|_{r=r_{m}}.
\end{equation}

By putting all together, we obtain that the derivative of $S_{WDW}$ with respect to $t_{0}$ for $t_{0}>t_{c}$ is given by
\begin{widetext}
\begin{equation}
\frac{dS_{WDW}}{dt_{0}}=2\mathcal{N}\left(\int_{r_{min}}^{r_{m}}dr\sqrt{-g}\mathcal{L}(r)-\mathcal{G}(r_{min})\right. + \left.\left.\sqrt{-\frac{g_{tt}}{g_{rr}}}\left(\sqrt{\sigma}\frac{g_{tt}'}{g_{tt}}-\frac{\sigma'}{\sqrt{\sigma}}\log\left(\frac{\sigma'\,l_{null}}{2\sigma\sqrt{-g_{rr}}}\right)\right)\right|_{r=r_{m}}\right).
\label{SWDWlate}
\end{equation}
\end{widetext}
Note that, as it should, the addition of the null counterterm has eliminated the arbitrary normalization constant $\alpha$, but the dependence on the arbitrary length scale $l_{null}$ remains. However, as we will explicitly show numerically in the next section, this term vanishes in the limit $t_{0}/t_{c}\rightarrow\infty$ for all the geometries that we consider. We also present in App. \ref{LateTime} an analytic proof of this vanishing that applies to metrics of the general form \eqref{PenroseMetric} that we are considering. We would like to stress that, even if we used an affine parametetrization in all of the previous calculations, the end result \eqref{SWDWlate} is invariant under reparametrizations of the null generators.

Given that both the MT and DK family of solutions are not analytical, we need to evaluate this integral numerically. We will show the result of said evaluation below. It follows from \eqref{CA} that the rate of change of the complexity of the TFD is related to $dS_{WDW}/dt_{0}$ by
\begin{eqnarray}
\frac{dC}{d\tau}=\frac{1}{2}\frac{dS_{WDW}}{dt_{0}}.
\end{eqnarray}

\subsection{Counterterms and the renormalization scheme}

From \eqref{SWDWlate} we can see that all the terms that diverge in the $r_{max}\rightarrow \infty$ limit were eliminated by the time derivative, rendering $dS_{WDW}/dt_{0}$ finite. This is because the $r=r_{max}$ surfaces only undergoes a time translation, and given that both families of solutions are static, any boundary integral evaluated at this surface will be independent of $t_{0}$. This is true for the counterterm actions \eqref{AnisotropicCounterterms} and \eqref{MagneticCounters}, which is why we did not included them in the evaluation of $S_{WDW}$. It should be noted though, that $S_{WDW}$ itself is a divergent quantity. The counterterms necessary to remove said divergences were computed in \cite{Akhavan:2019zax}, and latter applied in \cite{Omidi:2020oit} for a BTZ black hole. These counterterms don't modify the late time behavior of the complexity rate of change, thus we omitted them on the previous computation.

Usually, when computing thermodynamic quantities such as the free energy of the state, the gravitational action is evaluated in the exterior region. The boundary of this region is constituted by the surfaces at $r=r_{max}$ and $r=r_{h}$, but the counterterms vanish when evaluated at the horizon. However, as the WDW patch boundary includes the $r=r_{min}$ surfaces, the counterterms need to be evaluated there too. We explicitly checked numerically that, for any solution in both models, the contribution coming from this vanishes when the limit $r_{min}\rightarrow r_{s}$ is taken. This means that $S_{WDW}$, and as a consequence the complexity of the TFD state, is independent of the finite term determined by $C_{sch}$.

\subsection{Results for the MT model}
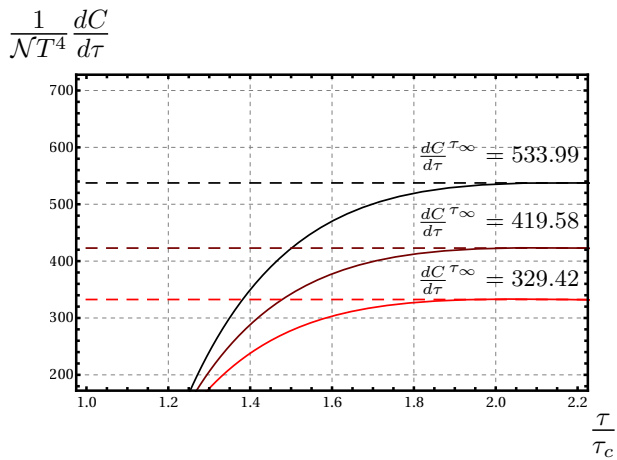
\begin{figure}[ht!]
 \centering
 \includegraphics[width=0.4\textwidth]{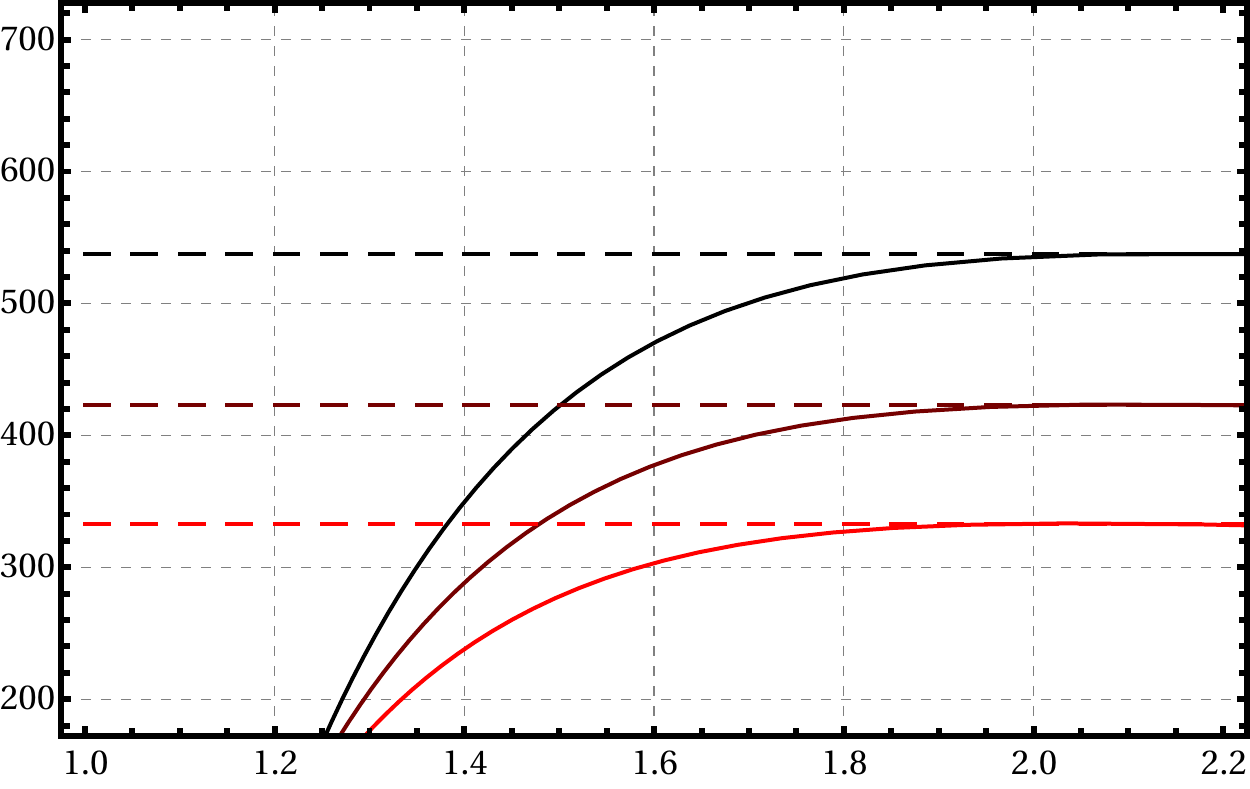}
 \put(-0,-10){\Large $\frac{\tau}{\tau_{c}}$}
 \put(-220,140){\Large $\frac{1}{\mathcal{N}T^{4}}\frac{dC}{d\tau}$}
\put(-65,48){$\frac{dC}{d\tau}^{\tau_\infty}=329.42$}
 \put(-65,70){$\frac{dC}{d\tau}^{\tau_\infty}=419.58$}
 \put(-65,95){$\frac{dC}{d\tau}^{\tau_\infty}=533.99$}
\caption{\small Rate of change of the complexity $dC/d\tau$ in units of $\mathcal{N}T^{4}$ as a function of $\tau$ for the MT model. Each curve corresponds to a different value for the anisotropic parameter, being $a/T=\{ 19.57, 41.14, 85.46 \}$ from bottom to top respectively. The horizontal dashed lines correspond to the late time behavior of each curve, with the precise values being $\frac{dC}{d\tau}^{\tau_\infty}=\{329.42, 419.58, 533.99\}$ respectively. For all cases we fixed $l_{null}=L=1$.}
\label{AnisotropicComplexity}
\end{figure}

\begin{figure}[ht!]
 \centering
 \includegraphics[width=0.4\textwidth]{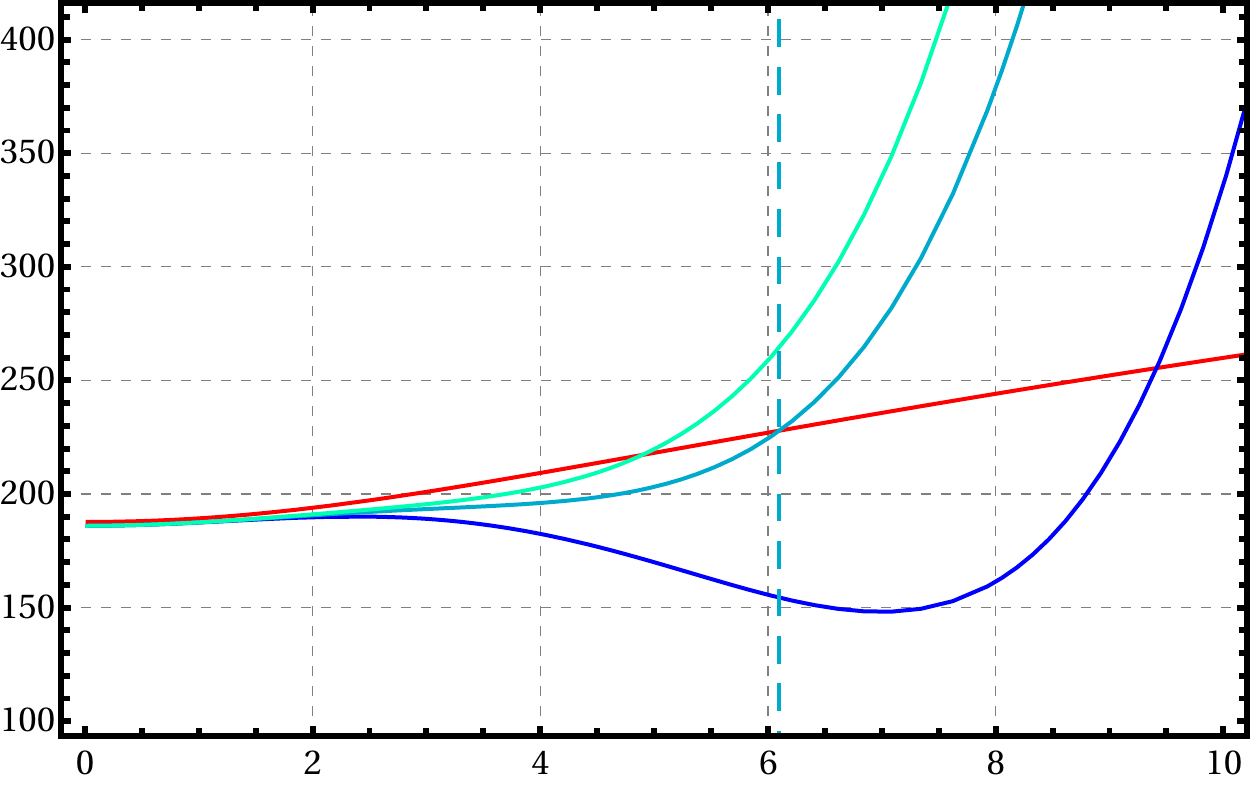}
 \put(-0,-10){\Large $\frac{a}{T}$}
\caption{\small Check of Lloyd's bound for the MT model as a function of $a/T$. The red curve (lower on the right side) corresponds to $\frac{dC}{d\tau}^{\tau_\infty}$, while the curves in hues of blue correspond to the energy of the TFD state $2E/\pi$ for $C_{sch}=\{-2,-3,-5\}$ from left to right respectively. All quantities are given in units of $\mathcal{N}T^{4}$.}
\label{AnisotropicLloyd}
\end{figure}

For the case of the MT model we have, after using the equations of motion, that
\begin{equation}
\mathcal{L}(r)=-8,
\end{equation}
\begin{equation}
\mathcal{G}(r)=\frac{r^{4}e^{-\frac{5}{4}\phi}(r(\mathcal{F}\mathcal{B})'+\mathcal{B}\mathcal{F}\left(8-3r\phi'\right))}{\sqrt{\mathcal{B}}},
\end{equation}
\begin{equation}
\sigma(r)=r^{6}e^{-\frac{5}{2}\phi}.
\end{equation}
With this expressions at hand, we can easily evaluate \eqref{SWDWlate} numerically, which allows us to compute the rate of change of the complexity of the TFD state as a function of $\tau$ for any value of the anisotropic parameter. We show the result of said evaluation in Fig. (\ref{AnisotropicComplexity}) for $l_{null}=L=1$ and the three values of the anisotropy $a/T=\{19.57, 41.14, 85.46\}$, displayed from bottom to top. This shows that the general effect of the anisotropy is to increase the value of $dC/d\tau$ for any given $\tau$. We can also see that, for any $a/T$, at $\tau$ shortly after $\tau_{c}$ the rate of change of the complexity decreases as time passes, reaches a minimum, and then it increases to a constant value, that we will denote as $\frac{dC}{d\tau}^{\tau_\infty}$, for late times $\tau\gg\tau_{c}$. We have placed dashed horizontal lines here in Fig. (\ref{AnisotropicComplexity}) to mark these asymptotic values, and plot them as a continuous function of $a/T$ displayed as a red line in Fig. (\ref{AnisotropicLloyd}). While the early time behavior can be modified by changing $l_{null}$, the late time behavior is independent of this arbitrary constant. It is also important to note that, for a given $\tau$, the rate of change of the complexity only depends on the dimensionless parameter $a/T$. We checked that this was the case numerically by varying $a$ and $T$ independently. We conclude from this that $dC/d\tau$ is independent of the energy scale $\mu$, and thus is unaffected by the conformal anomaly. 

\begin{figure}[ht!]
 \centering
 \includegraphics[width=0.4\textwidth]{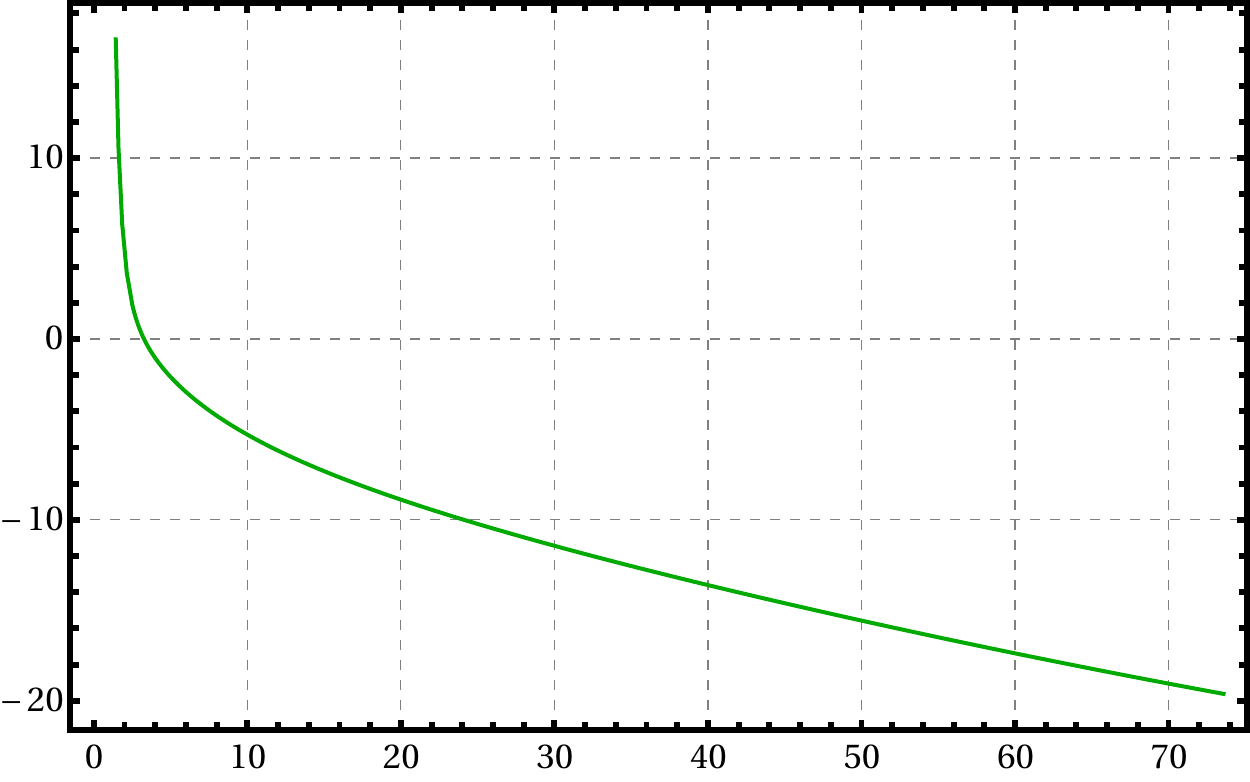}
 \put(-0,-10){\Large $\frac{a}{T}$}
 \put(-225,120){\Large $C_{sch}$}
\caption{\small Running of $C_{sch}$ with $a/T$ to keep the renormalization condition of saturating Lloyd's bound in the MT model.}
\label{AnisotropicScheme}
\end{figure}

As mentioned above, in Fig. (\ref{AnisotropicLloyd}) we show the late time behavior of the rate of change of the complexity $\frac{dC}{d\tau}^{\tau_\infty}$ as a function of $a/T$ (red curve), and compare it to the energy of the state in order to check if Lloyd's bound is verified at infinite time for the MT model, as this was the case for $a=0$. As explained in Sec. \ref{GravitySetup}, we have set the renormalization scale as $\mu=L=1$, but up to this point we still do not have a physical reason to fix the value of $C_{sch}$. Thus, in Fig. (\ref{AnisotropicLloyd}) we plot the energy of the state as a function of $a/T$ for $C_{sch}=\{-2,-3,-5\}$ shown by the curves from left to right, additional to the one representing $\frac{dC}{d\tau}^{\tau_\infty}$. We see that at any value of $a/T$ Lloyd's bound at infinite time can be satisfied, saturated, or violated depending on the value of $C_{sch}$, as we exemplify for $a/T=6.10$, shown by the vertical dashed line in Fig. (\ref{AnisotropicLloyd}). For this to be an actual bound regardless of how large the anisotropy is, it would be necessary to push $C_{sch}$ all the way to $\infty$, invalidating all physical calculations. Instead we suggest to chose the value of $C_{sch}$ that at each $a/T$ leads to the saturation of Lloyd's bound. We show this value of $C_{sch}$ as a function of $a/T$ in Fig. (\ref{AnisotropicScheme}). Physically, this amounts to use the saturation of Lloyd's bound at infinite time as a renormalization condition and let $C_{sch}$ run with $a/T$ to satisfy it.

\subsection{Results for the DK model}
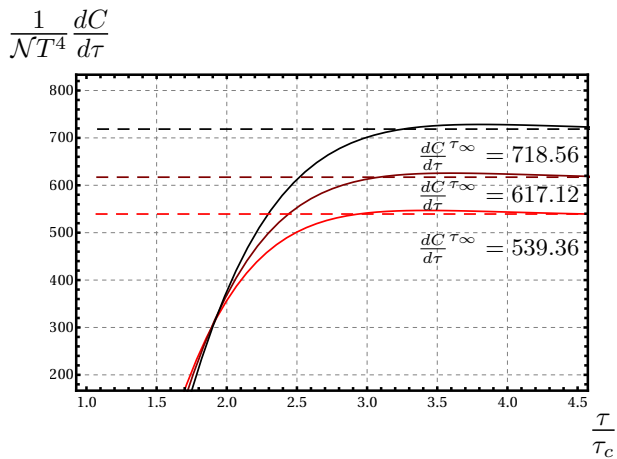
\begin{figure}[ht!]
 \centering
 \includegraphics[width=0.4\textwidth]{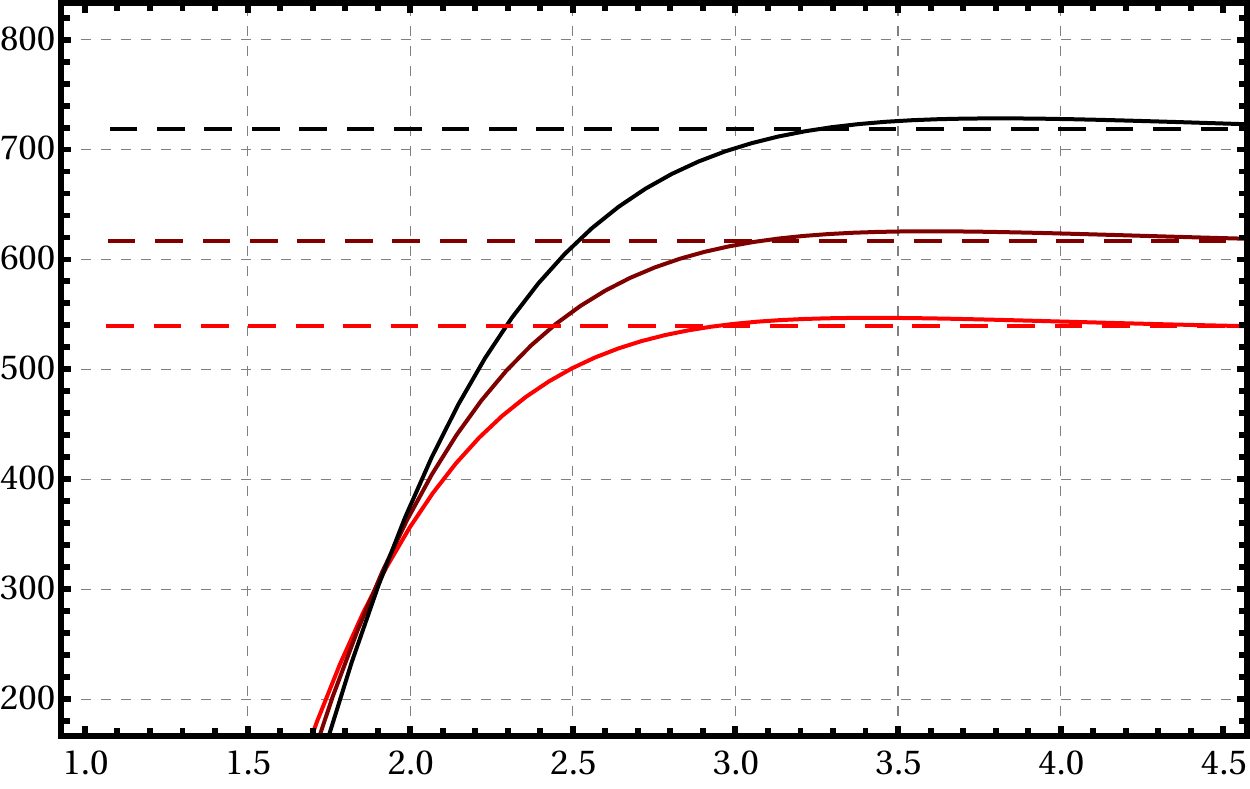}
 \put(-0,-10){\Large $\frac{\tau}{\tau_{c}}$}
 \put(-220,140){\Large $\frac{1}{\mathcal{N}T^{4}}\frac{dC}{d\tau}$}
 \put(-65,60){$\frac{dC}{d\tau}^{\tau_\infty}=539.36$}
 \put(-65,80){$\frac{dC}{d\tau}^{\tau_\infty}=617.12$}
 \put(-65,95){$\frac{dC}{d\tau}^{\tau_\infty}=718.56$}
\caption{\small Rate of change of the complexity $dC/d\tau$ in units of $\mathcal{N}T^{4}$ as a function of $\tau$ for the DK model. Each curve corresponds to a different magnetic field intensity, being $b/T^{2}=\{40.59, 47.56, 56.62 \}$ from bottom to top respectively. The horizontal dashed lines correspond to the late time behavior of each curve, with the precise values being $\frac{dC}{d\tau}^{\tau_\infty}=\{539.36, 617.12, 718.56\}$ respectively. For all cases we fixed $l_{null}=L=1$.}
\label{MagneticComplexity}
\end{figure}

For the case of the DK model we have, after using the equations of motion, that
\begin{equation}
\mathcal{L}(r)=-8-\frac{4b^{2}}{3V^{2}},
\end{equation}
\begin{equation}
\mathcal{G}(r)=\frac{UWV'+(UWV)'}{\sqrt{W}},
\end{equation}
\begin{equation}
\sigma(r)=WV^{2}.
\end{equation}
Employing these expressions we can evaluate \eqref{SWDWlate} numerically, which allows us to compute the complexity rate of change on the TFD state as a function of $\tau$ for any magnetic field intensity. In Fig. (\ref{MagneticComplexity}) we show the result of this evaluation for $l_{null}=L=1$ with three different values of the magnetic  field intensity being, from bottom to top, $b/T^{2}=\{40.59, 47.56, 56.62 \}$ respectively. We see that the magnetic field has the general effect of increasing the value of $dC/d\tau$ for any given $\tau$. The behaviour common to the three plots in Fig. (\ref{MagneticComplexity}) indicates that for any $b/T^{2}$, at $\tau$ shortly past $\tau_{c}$ the  complexity rate of change is an increasing function of time until it reaches a maximum, after which it becomes a decreasing function that for $\tau\gg\tau_{c}$ tends to a constant value, which we denote as $\frac{dC}{d\tau}^{\tau_\infty}$ and indicate in the three cases of Fig. (\ref{MagneticComplexity}) as horizontal dashed lines. While the early time behavior can be modified by varying $l_{null}$, the late time behavior is unaffected by it. We also explicitly checked that, for a given $\tau$, the rate of change of the complexity only depends on the dimensionless parameter $b/T^{2}$. In practice, we achieved this by varying $b$ and $T$ independently. We conclude from this that $dC/d\tau$ is independent of the energy scale $\mu$, proving that this quantity is unaffected by the conformal anomaly. 

\begin{figure}[ht!]
 \centering
 \includegraphics[width=0.4\textwidth]{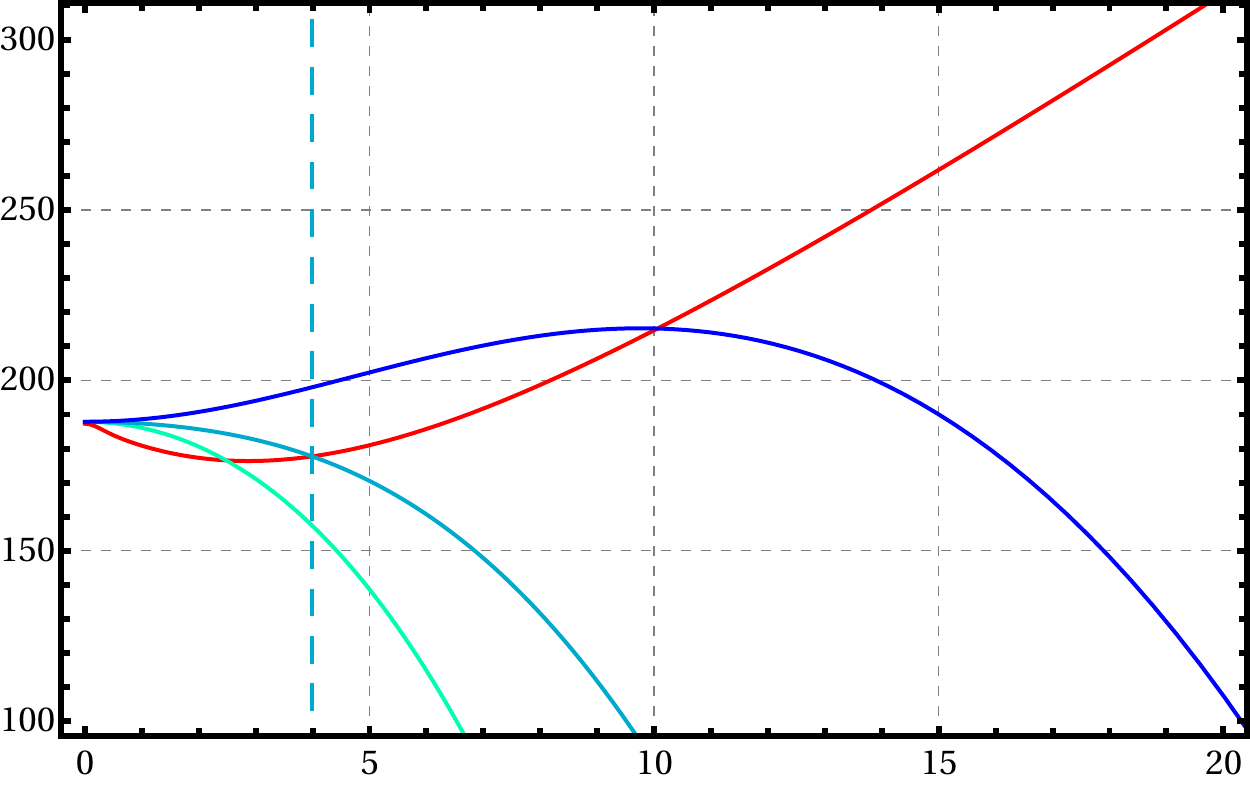}
 \put(-0,-10){\Large $\frac{b}{T^{2}}$}
\caption{\small Check of Lloyd's bound for the DK model as a function of $b/T^{2}$. The red curve (higher on the right) corresponds to $\frac{dC}{d\tau}^{\tau_\infty}$, while the curves in hues of blue correspond to the energy of the TFD state $2E/\pi$ for $C_{sch}=\{0,1,2\}$ from left to right respectively. All quantities are given in units of $\mathcal{N}T^{4}$.}
\label{MagneticLloyd}
\end{figure}

\begin{figure}[ht!]
 \centering
 \includegraphics[width=0.4\textwidth]{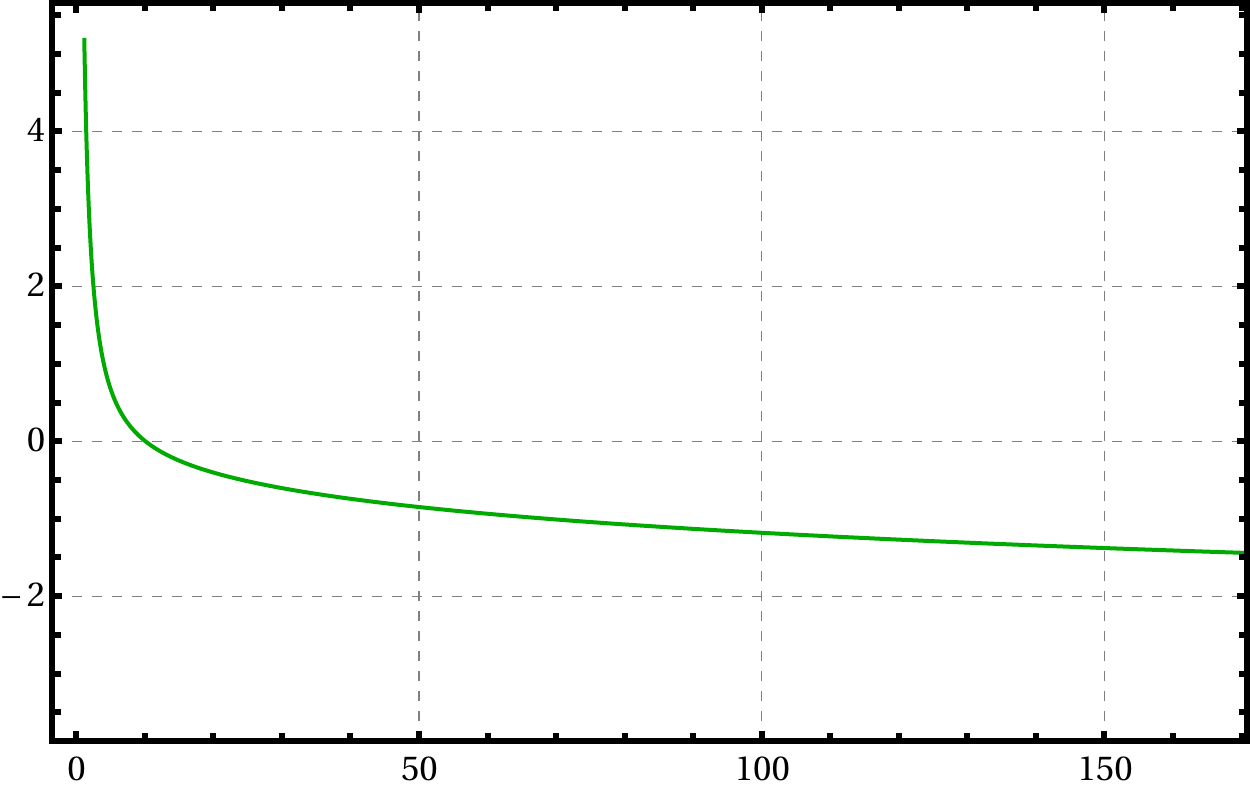}
 \put(-0,-10){\Large $\frac{b}{T^{2}}$}
 \put(-225,120){\Large $C_{sch}$}
\caption{\small Running of $C_{sch}$ with $b/T^{2}$ to maintain the renormalization condition of the saturation of Lloyd's bound for the DK model.}
\label{MagneticScheme}
\end{figure}

We show $\frac{dC}{d\tau}^{\tau_\infty}$ as a function of $b/T^{2}$ (red curve) in Fig. (\ref{MagneticLloyd}), and compare it to the energy of the state in order to check if Lloyd's bound is verified in the DK model at infinite time, as this was the case for $b=0$. We have fixed the renormalization scale $\mu=L=1$ but, as explained in Sec. \ref{GravitySetup}, up to this point of the analysis we do not have a physical reason to fix $C_{sch}$. Hence, in Fig. (\ref{MagneticLloyd}) we show the plots of the energy of the state as a function of $b/T^{2}$ for $C_{sch}=\{0,1,2\}$, from left to right. Just as it was the case for the MT model, and as we exemplify at $b/T^{2}=3.99$, Lloyd's bound at infinite time can be satisfied, saturated, or violated for any $b/T^{2}\neq 0$ depending on the value of $C_{sch}$, and the only way to respect Lloyd's bound for any $b/T^{2}$ is by setting $C_{sch}=\infty$. Once again we decide to adjust $C_{sch}$ at each value of $b/T^{2}$ so that Lloyd's bound is saturated. As before, the latter corresponds to the physical proposal of letting $C_{sch}$ run with $b/T^{2}$, as we show in Fig. (\ref{MagneticScheme}), to maintain the renormalization condition of the saturation of Lloyd's bound at infinite time.

\section{Discussion}
\label{Discussion}
In this paper we employed holographic methods to determine the effect that the conformal anomaly has on the computational complexity. To this end, we extended the study presented in \cite{HosseiniMansoori:2018gdu} to arbitrary anisotropies and magnetic field intensities by considering the numerical families of solutions for both the MT and DK models. While ours coincide with previous results where applicable, we were able to derive many novel results.

Our first main result is that the rate of change of the complexity is unaffected by the conformal anomaly present in both models. The first of two reason why this is the case is that $(1/T^{4})dC/d\tau$, which is a dimensionless quantity, only depends on $a$ and $T$ through the dimensionless ratio $a/T$, proving that such derivative is independent of the energy scale $\mu$, which we fixed to the unity throughout this paper. The second reason is that the counterterm action, which contains the scheme dependent coefficient $C_{sch}$, does not contribute to the derivative with respect to $\tau$, as the integral over the boundary regulator is constant and the integral near the singularity vanishes as the regulator is removed. 

The other main result is in regards of Lloyd's bound. It is expected, when using the CA prescription, that Lloyd's bound will be violated for any finite time, only to be saturated when an infinite amount of time have passed. Knowing that this was the case for $a=0$ and $b=0$, we studied the validity of this result for arbitrary values of the anisotropic parameter and the magnetic field. As explained in Sec. \ref{Intro}, the energy of the system, which appears on the right hand side of Lloyd's bound \eqref{Lloyd}, depends on both the energy scale $\mu$ and the coefficient $C_{sch}$. However as just stated, $dC/d\tau$, which appears on the right hand side of \eqref{Lloyd}, does not depend on any of this quantities. While at first sight this could be interpreted as an inconsistency for Lloyd's bound when a conformal anomaly is present, that is not what it is. The role of $C_{sch}$ is to keep physical quantities scheme independent, and in particular for the the energy, this means to absorb any modification that it could suffer when the value of $\mu$ is changed, once of course a renormalization condition has been imposed. Thus, the results above demonstrate that, if so desired, the saturation of Lloyd's bound at infinite time can be used as a renormalization condition and let $C_{sch}$ be adjusted to satisfy it for any given $\{\mu,a,T\}$ or $\{\mu,b,T\}$, depending on the model.

We would like to stress the importance of $dC/d\tau$ being independent of both $\mu$ and $C_{sch}$. If it had turned out to depend on only one, there would have been no way to cancel any arbitrary change in the other, making it a scheme dependent quantity, and therefore, questioning the validity of the CA conjecture. Clearly the other acceptable option was for $dC/d\tau$ to depend on both, $\mu$ and $C_{sch}$, so we consider interesting to have disproved this possibility.

It is also important to note that, to our knowledge, a renormalization condition has not been proposed previously for neither the MT or the DK model.  

\section*{Acknowledgments}
We acknowledge partial financial support from PAPIITIN113618, UNAM. DA, CD and YO are supported by CONACYT Ph.D. grants.
\appendix
\section{Appendix}

\subsection{Interior solutions}
\subsubsection{MT model}
\label{AppAnisotropic}
In this appendix we explain the integration procedure needed to obtain the interior solutions for the MT model. The equations of motion that determine the metric functions and the dilaton and axion fields come from the variation of the 5-dimensional action \eqref{AnisotropicAction}. After the substitution of the ansatz \eqref{AnisotropicAnsatz}, this equations can be manipulated to write $\mathcal{F}$ and $\mathcal{B}$ in terms of $\phi$ as
\begin{equation}
\mathcal{F}=\frac{e^{-\frac{1}{2}\phi}(a^{2}e^{\frac{7}{2}\phi}(r\phi'-4)+16 r^{3}\phi')}{4r^{3}(\phi'+r\phi'')},
\label{EOMF}
\end{equation}
\begin{equation}
\frac{\mathcal{B}'}{\mathcal{B}}=\frac{-16\phi'+9r\phi'^{2}-20r\phi''}{24-10r\phi'},
\label{EOMB}
\end{equation}
while the dilaton itself is given by the third order differential equation
\begin{equation}
\begin{split}
0=&8r\left(r^{2}\left(11a^{2}e^{\frac{7\phi}{2}}+96r^{2}\right)\phi''^{2}+12a^{2}re^{\frac{7\phi}{2}}\phi''' \right) \\ & + 8 r\left(12a^{2} e^{\frac{7\phi}{2}}\phi''\right)+\left(352r^{5}-13a^{2}r^{3}e^{\frac{7\phi}{2}}\right)\phi'^{4}\\&+2r\phi'^2\left(r^{2}\left(23a^{2}e^{\frac{7\phi}{2}}-880r^{2}\right)\phi'' \right) \\ & + 2 r \phi'^{2} \left( 5r^{3}\phi'''\left(a^{2}e^{\frac{7\phi}{2}}+16r^{2}\right)-12 a^{2}e^{\frac{7\phi}{2}}+960r^{2}\right)\\&+\phi'^3\left(8r^{2}\left(7a^{2}e^{\frac{7\phi}{2}}-200r^{2}\right) \right) \\& + \phi'^{3} \left( \left(352r^{6}-13a^{2}r^{4}e^{\frac{7\phi}{2}}\right)\phi''\right) \\ &-2\phi'\left(32r^{2}\left(a^{2}e^{\frac{7\phi}{2}}-30r^{2}\right)\phi''\right.\\&\left.+15r^{4}\left(a^{2}e^{\frac{7\phi}{2}}+16r^{2}\right)\phi''^{2}+32r^{3}\phi'''\left(a^{2}e^{\frac{7\phi}{2}}+6r^{2}\right) \right) \\ & + 15r^4\left(48a^{2}e^{\frac{7\phi}{2}}\right).
\end{split}
\label{EOMPhi}
\end{equation}

As explained in \cite{Mateos:2011tv}, it is possible to eliminate the anisotropic parameter $a$ from \eqref{EOMPhi} altogether by shifting the dilaton field to
\begin{equation}
\tilde{\phi}=\phi+\frac{4}{7}\log{a},
\label{PhiTilde}
\end{equation}
and thus solve directly for $\tilde{\phi}$. This allows us to easily impose the boundary condition $\phi_{bdry}=0$ by defining that the anisotropy corresponding to a given solution is given by
\begin{equation}
a=e^{\frac{7}{4}\tilde{\phi}_{bdry}}.
\label{aPhiTilde}
\end{equation}
Thus, solutions with different $\tilde{\phi}_{bdry}$ correspond to solutions with different anisotropies.

Given that the equation for $\tilde{\phi}$ is highly non-linear, it is necessary to resort to numerical methods to obtain a solution for an arbitrary $a/T$. The first step is to expand \eqref{EOMPhi} in a power series of $r$ around the horizon
\begin{equation}
\tilde{\phi}=\tilde{\phi}_{h}+\sum_{i=1}^{\infty}\tilde{\phi}_{i}(r-r_{h})^{i}.
\label{HorizonPhi}
\end{equation}
By solving order by order it is possible to write any undetermined coefficient\footnote{It is also necessary to impose that $\mathcal{F}(r_{h})=0$ by means of (A1).} $\phi_{i}$ in terms of $\tilde{\phi}_{h}$. Then \eqref{HorizonPhi} is used to provide initial data for the numerical integration, starting from $r=r_{h}+\epsilon$ with $\epsilon\ll r_{h}$ all the way to the boundary at $r=\infty$ in the case of the exterior solutions, and from $r=r_{h}-\epsilon$ towards, and down to, the singularity at $r=0$ in the case of the interior solutions. After this we can obtain the anisotropy by substitution in \eqref{aPhiTilde} and finally $\phi$ with the appropriate boundary condition using \eqref{PhiTilde}.

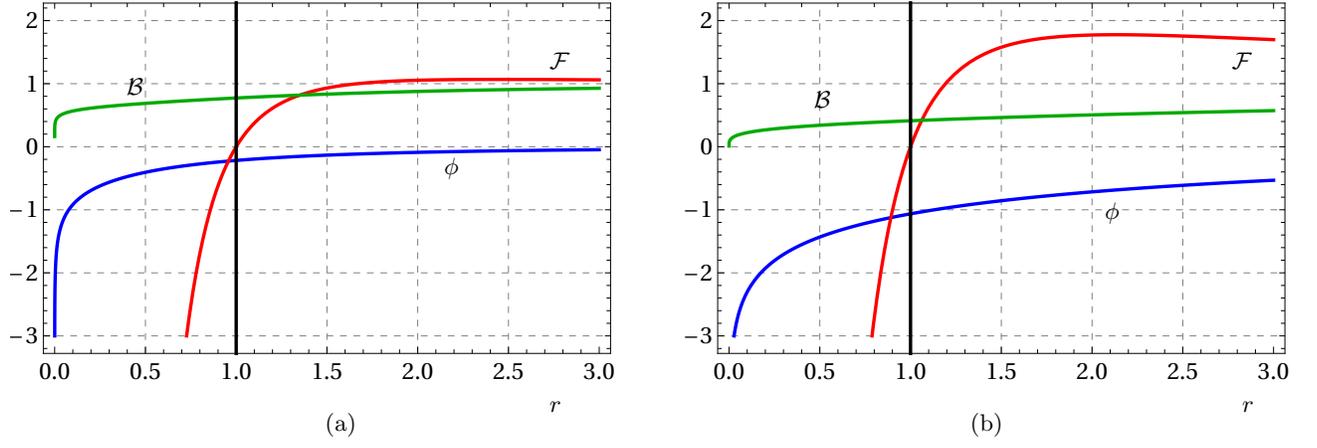
\begin{figure*}
\begin{center}
\begin{tabular}{cc}
\includegraphics[width=0.45\textwidth]{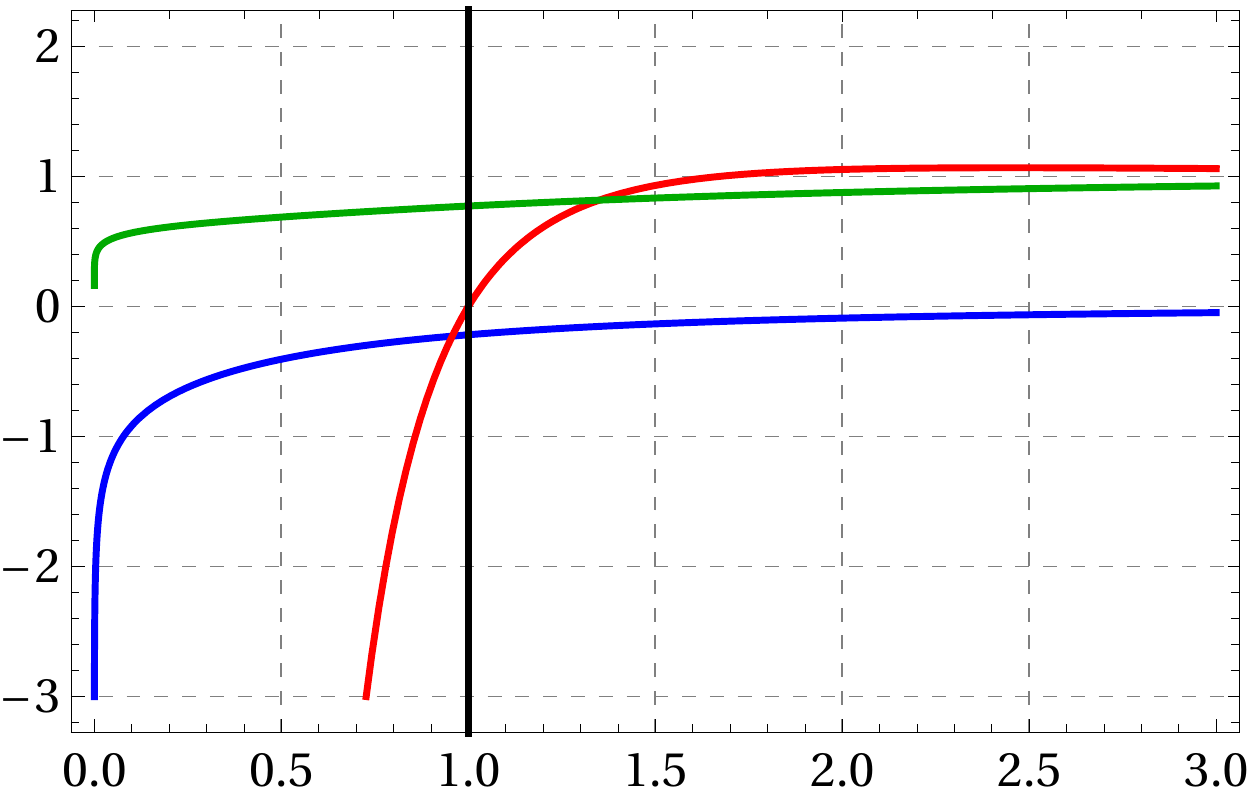} 
\qquad\qquad & 
\includegraphics[width=0.45\textwidth]{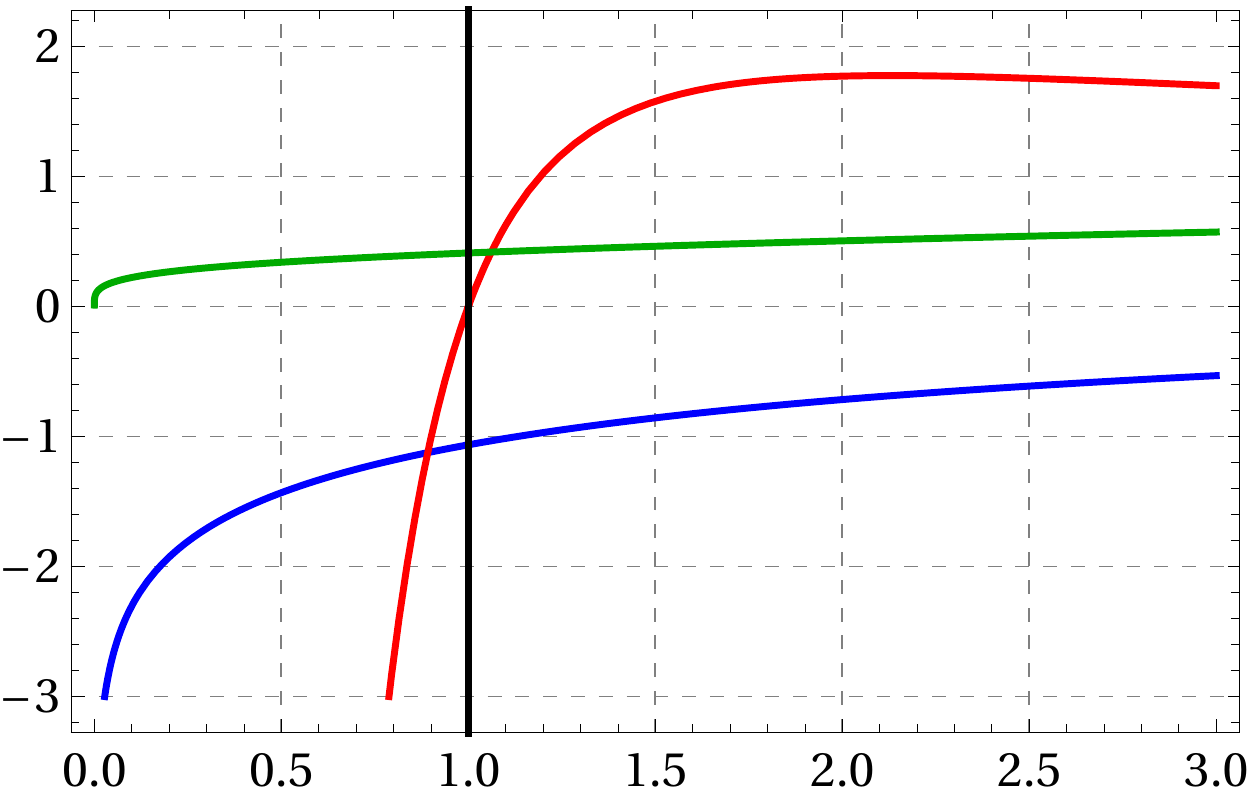}
\qquad
   \put(-280,-10){$r$}
   \put(-280,120){$\mathcal{F}$}
   \put(-440,110){$\mathcal{B}$}
   \put(-320,80){$\phi$}
   \put(-18,-10){$r$}
   \put(-22,120){$\mathcal{F}$}
   \put(-180,105){$\mathcal{B}$}
   \put(-70,63){$\phi$}
 \\
(a) & (b)\\
\end{tabular}
\end{center}
\caption{\small Metric functions and dilaton for the MT model as functions of $r$ for (a) $a/T=4.41$ and (b) $a/T=24.91$. In both figures the horizon, located at $r_{h}=1$, is denoted as a black vertical line.}
\label{AnisotropicPlot}
\end{figure*}

With the solution for $\phi$ in the interior and the exterior at hand, we can obtain $\mathcal{F}$ by simply substituting in \eqref{EOMF} as this is an algebraic relation. Incidentally, this always results in an $\mathcal{F}$ satisfying the boundary condition $\mathcal{F}_{bdry}=1$. However, given that the relation between $\mathcal{B}$ and $\phi$ is differential, it is necessary to perform again a numerical integration. In order to do this we expand \eqref{EOMB} in a power series of $r$ around $r_{h}$ using \eqref{HorizonPhi} and
\begin{eqnarray}
\mathcal{B}=\mathcal{B}_{h}+\sum_{i=1}^{\infty}\mathcal{B}_{i}(r-r_{h})^{i}.
\end{eqnarray}
By solving \eqref{EOMB} order by order, we can write any undetermined coefficient in terms of $\mathcal{B}_{h}$ and $\phi_{h}$. Then, using the result of this as initial data,  we can integrate \eqref{EOMB} numerically from $r=r_{h}+\epsilon$ to the boundary at $r=\infty$ for the exterior solutions, and from $r=r_{h}-\epsilon$ to the singularity at $r=0$ for the interior solutions. Given that we want the spacetime to asymptote $AdS_{5}$ at the boundary, the last step is to use the symmetry of \eqref{EOMB} and redefine
\begin{equation}
\mathcal{B}\rightarrow\frac{\mathcal{B}}{\mathcal{B}_{bdry}},
\end{equation}
ensuring in this way that $\mathcal{B}$ goes to unity at the boundary. Note that this rescaling needs to be done consistently for both the exterior and interior solutions. In Fig. (\ref{AnisotropicPlot}) we show the metric functions for small ($a/T=4.41$) and large ($a/T=24.91$) anisotropies in the interior and exterior regions. 

\subsubsection{DK model}
\label{AppMagnetic}
In this appendix we show the integration procedure needed to obtain the interior solutions for the DK model. The equations of motion that determine the metric functions and the Maxwell field come from the variation of the 5-dimensional action \eqref{MagneticAction}. After the substitution of the ansatz \eqref{MagneticAnsatz}, Maxwell equations are automatically satisfied and Einstein equations can be manipulated into
\begin{equation}
\begin{split}
0= & 2W^{2}\left(4b^{2}+V\left(U'V'+UV''\right)\right)-\\ & VW\left(2V\left(U'W'+UW''\right)+UV'W'\right)+UV^{2}W'^{2}, \\
0= & -2W^{2}\left(V'^{2}-2VV''\right)+2V^{2}WW''-V^{2}W'^{2}, \\
0= & W\left(8b^{2}-6V^{2}\left(U''-8\right)-6VU'V'\right)-3V^{2}U'W', \\
0= & W\left(4b^{2}+2VU'V'+UV'^{2}-24V^{2}\right) \\ & +VW'\left(VU'+2UV'\right).
\end{split}
\label{EOMMagnetic}
\end{equation}
Given that the equations are singular at $r_{h}$, in order to solve numerically we first expand them in powers of $r$ around $r_{h}$ using
\begin{eqnarray}
&& U=6r_{h}(r-r_{h})+\sum_{i=2}^{\infty}U_{i}(r-r_{h})^{i},
\cr
&& V=V_{0}+\sum_{i=1}^{\infty}V_{i}(r-r_{h})^{i},
\cr
&& W=3r_{h}^{2}+\sum_{i=1}^{\infty}W_{i}(r-r_{h})^{i}.
\label{HorizonMagnetic}
\end{eqnarray}
This behavior near the horizon allows the family of solutions to easily interpolate between the D3-black brane and BTZ$\times \mathbb{R}^{2}$ by changing the value of $b/V_{0}$ from $0$ to $\sqrt{3}$. Additionally, this also ensures that the temperature of every member of the family is given by $T=3 r_{h}/2\pi$. However, it is important to note that the coordinate $r$ here is not the usual radial coordinate of the black brane solution $\tilde{r}$. Instead, the relation between the two is given by
\begin{eqnarray}
\tilde{r}=r+\frac{r_{h}}{2},
\end{eqnarray}
which means that for vanishing magnetic field the singularity is located at $r=-r_{h}/2$ and not at $r=0$.

By plugging \eqref{HorizonMagnetic} into \eqref{EOMMagnetic} we can solve for any of the undetermined coefficients in terms of $b/V_{0}$, and then use this to provide initial data for the numerical integration. This is performed from $r=r_{h}+\epsilon$ to the boundary at $r=\infty$ for the exterior solutions, and from $r=r_{h}-\epsilon$ to the singularity at $r=r_{s}$ for the interior solutions, with $\epsilon\ll r_{h}$ in both cases. This procedure will give solutions which boundary behavior goes like
\begin{equation}
V\sim V_{bdry}r^{2}, \qquad W\sim W_{bdry}r^{2}, \qquad U\sim r^{2}.
\end{equation}
However, we can exploit the symmetries of the equations of motion \eqref{EOMMagnetic} and rescale them as
\begin{eqnarray}
V\rightarrow \frac{V}{V_{bdry}}, \qquad W\rightarrow \frac{W}{W_{bdry}}, \qquad b\rightarrow \frac{b}{V_{bdry}},
\end{eqnarray}
which in turn gives the desired $AdS_{5}$ behavior at the boundary. Once again, this rescaling needs to be done consistently for both the exterior and interior solutions. Notice that in this occasion, it is necessary to simultaneously scale the value of $b$ to preserve the solution.

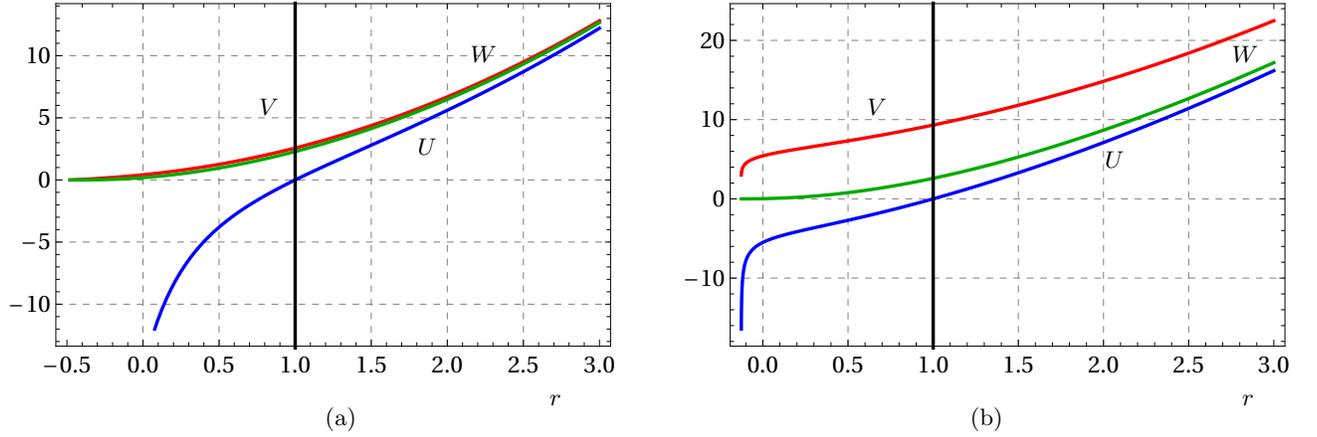
\begin{figure*}
\begin{center}
\begin{tabular}{cc}
\includegraphics[width=0.45\textwidth]{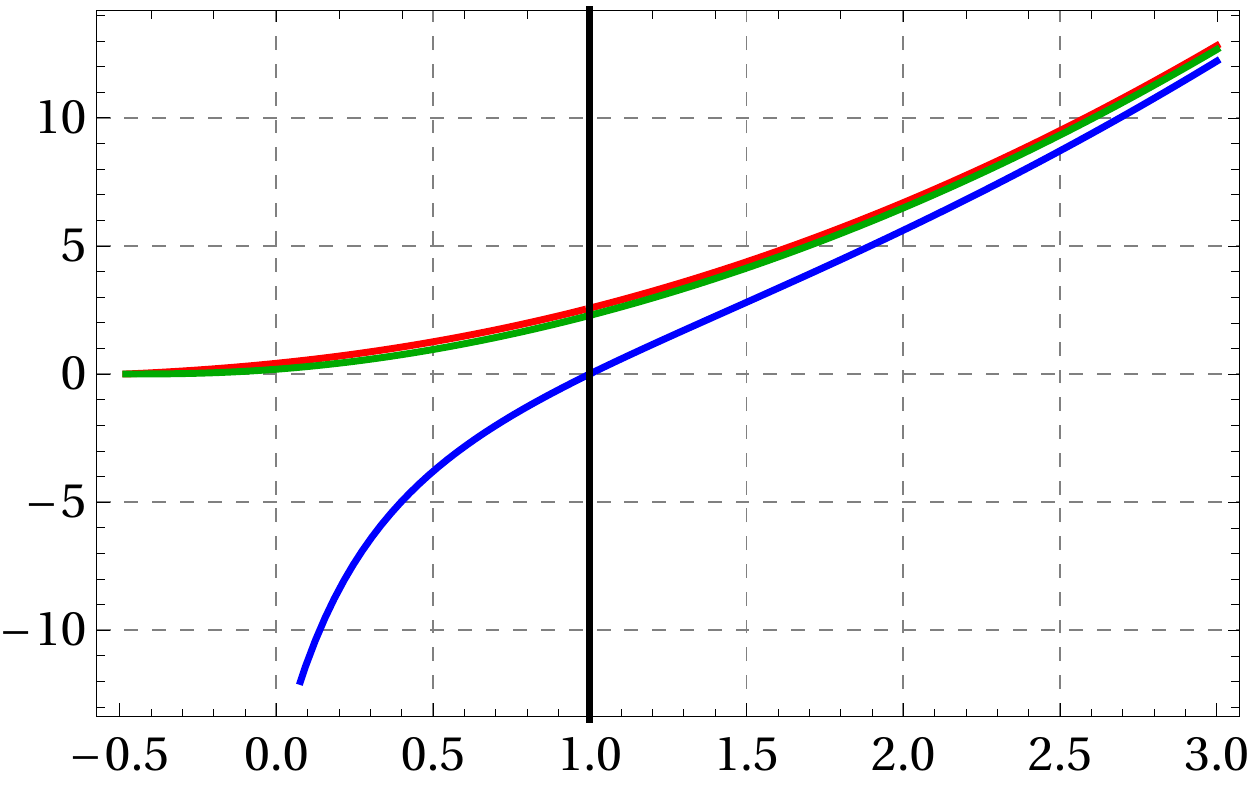} 
\qquad\qquad & 
\includegraphics[width=0.45\textwidth]{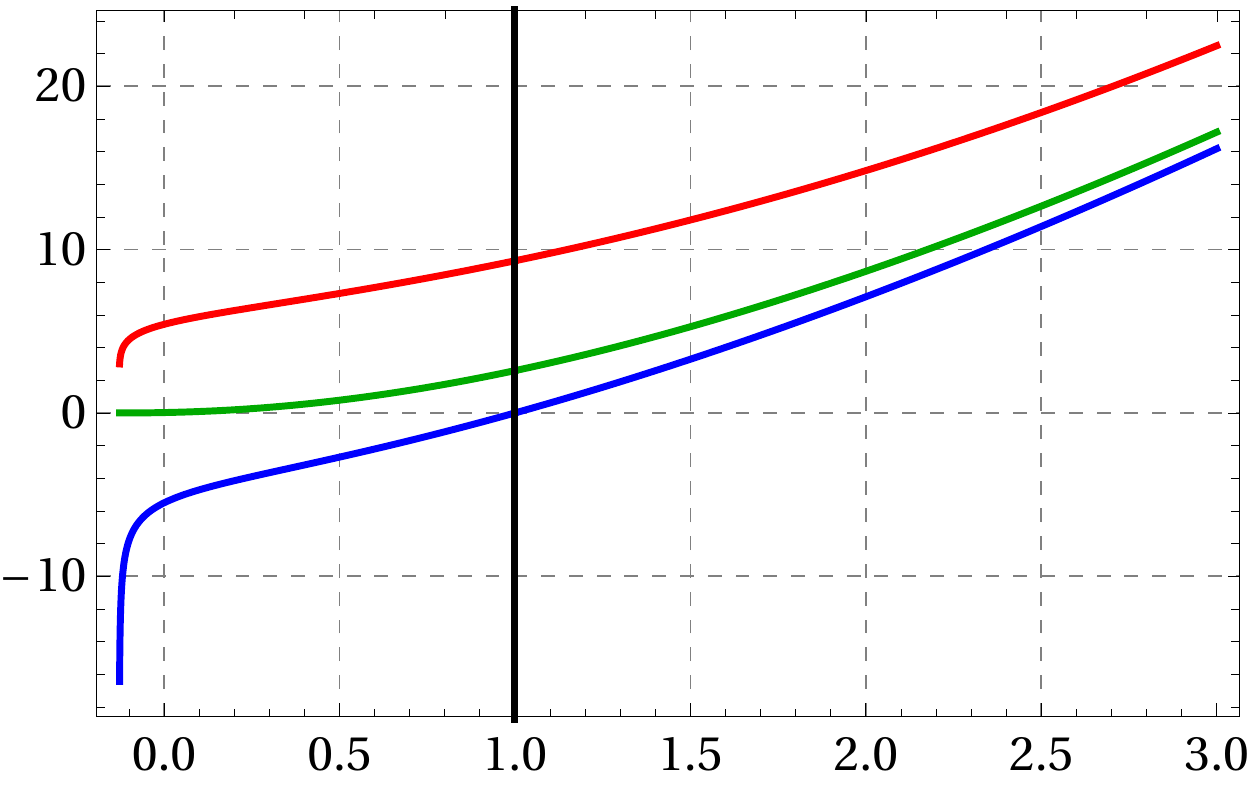}
\qquad
   \put(-280,-10){$r$}
   \put(-310,120){$W$}
   \put(-390,100){$V$}
   \put(-330,85){$U$}
   \put(-18,-10){$r$}
   \put(-22,120){$W$}
   \put(-160,100){$V$}
   \put(-70,80){$U$}
 \\
(a) & (b)\\
\end{tabular}
\end{center}
\caption{\small Metric functions for the DK model as functions of $r$ for (a) $b/T^{2}=5.85$ and (b) $b/T^{2}=56.62$. In both figures the horizon, located at $r_{h}=1$, is denoted as a black vertical line.}
\label{MagneticPlot}
\end{figure*}

One important thing to note is that the position of the singularity is not fixed at $r_{s}=-r_{h}/2$ for every member of the family of solutions, but only for $b/T^{2}=0$. Instead, the location of the singularity in the $r$-coordinate is a function of the magnetic field intensity. By $r_{s}$ we mean the radius at which the curvature scalar
\begin{eqnarray}
R_{\mu\nu\alpha\beta}R^{\mu\nu\alpha\beta},
\end{eqnarray}
goes to infinity. We show the metric functions for small ($b/T^{2}=5.85$) and large ($b/T^{2}=56.62$) magnetic field intensities in the interior and exterior regions in Fig. (\ref{MagneticPlot}).

\subsection{Late time behavior of the complexity rate of change}
\label{LateTime}

In this appendix we explicitly write the expression for the late time behavior of the complexity rate of change \eqref{SWDWlate} and show that the dependence on $l_{null}$ vanishes in the limit $t_{0}\rightarrow \infty$. In this limit $r_{m}$ goes to $r_{h}$, hence \eqref{SWDWlate} takes the form
\begin{widetext}
\begin{equation}
\lim_{t_{0}\rightarrow\infty}\frac{dS_{WDW}}{dt_{0}}= 2\mathcal{N}\left(\int_{r_{min}}^{r_{h}}dr\sqrt{-g}\mathcal{L}(r)-\mathcal{G}(r_{min})\right)+2\mathcal{N}\lim_{r_{m}\rightarrow r_{h}}\left(\left.\sqrt{-\frac{g_{tt}}{g_{rr}}}\left(\sqrt{\sigma}\frac{g_{tt}'}{g_{tt}}-\frac{\sigma'}{\sqrt{\sigma}}\log\left(\frac{\sigma'\,l_{null}}{2\sigma\sqrt{-g_{rr}}}\right)\right)\right|_{r=r_{m}}\right).
\end{equation}
\end{widetext}
To evaluate the limit in the second term we use the fact that $g_{tt}$ has a zero at $r_{h}$, $g_{rr}$ has a simple pole there, while the other metric functions are regular at the horizon. This means that to leading order
\begin{equation}
\sqrt{-\frac{g_{tt}}{g_{rr}}}\sim r-r_{h}, \quad \frac{\sigma'}{\sqrt{\sigma}}\sim (r-r_{h})^{0}, \quad \frac{\sigma'}{\sigma}\sim (r-r_{h})^{0},
\end{equation}
hence
\begin{equation}
\log\left(\frac{\sigma'\,l_{null}}{2\sigma\sqrt{-g_{rr}}}\right)\sim \log(r-r_{h}),
\end{equation}
which in turn means that
\begin{equation}
\lim_{r_{m}\rightarrow r_{h}}\sqrt{-\frac{g_{tt}}{g_{rr}}}\frac{\sigma'}{\sqrt{\sigma}}\log\left(\frac{\sigma'\,l_{null}}{2\sigma\sqrt{-g_{rr}}}\right)=0.
\end{equation}

This shows that $l_{null}$ does not affect the late time behavior of the rate of change of the complexity, as the final expression is

\begin{widetext}
\begin{equation}
\lim_{t_{0}\rightarrow\infty}\frac{dS_{WDW}}{dt_{0}}= 2\mathcal{N}\left(\int_{r_{min}}^{r_{h}}dr\sqrt{-g}\mathcal{L}(r)-\mathcal{G}(r_{min})+\left.\left(g_{tt}'\sqrt{-\frac{\sigma}{g_{rr}g_{tt}}}\right)\right|_{r=r_{h}}\right).
\end{equation}
\end{widetext}

\newpage
\bibliography{Complexity.bib}

\begin{thebibliography}{10}

\bibitem{Maldacena:1997re}
Juan~Martin Maldacena.
\newblock {The Large N limit of superconformal field theories and
  supergravity}.
\newblock {\em Int. J. Theor. Phys.}, 38:1113--1133, 1999.
\newblock [Adv. Theor. Math. Phys.2,231(1998)].

\bibitem{Gubser:1998bc}
S.~S. Gubser, Igor~R. Klebanov, and Alexander~M. Polyakov.
\newblock {Gauge theory correlators from noncritical string theory}.
\newblock {\em Phys. Lett. B}, 428:105--114, 1998.

\bibitem{Witten:1998qj}
Edward Witten.
\newblock {Anti-de Sitter space and holography}.
\newblock {\em Adv. Theor. Math. Phys.}, 2:253--291, 1998.

\bibitem{Ryu:2006bv}
Shinsei Ryu and Tadashi Takayanagi.
\newblock {Holographic derivation of entanglement entropy from AdS/CFT}.
\newblock {\em Phys. Rev. Lett.}, 96:181602, 2006.

\bibitem{Hubeny:2007xt}
Veronika~E. Hubeny, Mukund Rangamani, and Tadashi Takayanagi.
\newblock {A Covariant holographic entanglement entropy proposal}.
\newblock {\em JHEP}, 07:062, 2007.

\bibitem{Takayanagi:2017knl}
Tadashi Takayanagi and Koji Umemoto.
\newblock {Entanglement of purification through holographic duality}.
\newblock {\em Nature Phys.}, 14(6):573--577, 2018.

\bibitem{Penington:2019npb}
Geoffrey Penington.
\newblock {Entanglement Wedge Reconstruction and the Information Paradox}.
\newblock {\em JHEP}, 09:002, 2020.

\bibitem{Almheiri:2019hni}
Ahmed Almheiri, Raghu Mahajan, Juan Maldacena, and Ying Zhao.
\newblock {The Page curve of Hawking radiation from semiclassical geometry}.
\newblock {\em JHEP}, 03:149, 2020.

\bibitem{Almheiri:2019yqk}
Ahmed Almheiri, Raghu Mahajan, and Juan Maldacena.
\newblock {Islands outside the horizon}.
\newblock 10 2019.

\bibitem{Hartnoll:2020fhc}
Sean~A. Hartnoll, Gary~T. Horowitz, Jorrit Kruthoff, and Jorge~E. Santos.
\newblock {Diving into a holographic superconductor}.
\newblock {\em SciPost Phys.}, 10:009, 2021.

\bibitem{Hartman:2013qma}
Thomas Hartman and Juan Maldacena.
\newblock {Time Evolution of Entanglement Entropy from Black Hole Interiors}.
\newblock {\em JHEP}, 05:014, 2013.

\bibitem{Susskind:2014rva}
Leonard Susskind.
\newblock {Computational Complexity and Black Hole Horizons}.
\newblock {\em Fortsch. Phys.}, 64:24--43, 2016.
\newblock [Addendum: Fortsch.Phys. 64, 44--48 (2016)].

\bibitem{Stanford:2014jda}
Douglas Stanford and Leonard Susskind.
\newblock {Complexity and Shock Wave Geometries}.
\newblock {\em Phys. Rev. D}, 90(12):126007, 2014.

\bibitem{Brown:2015bva}
Adam~R. Brown, Daniel~A. Roberts, Leonard Susskind, Brian Swingle, and Ying
  Zhao.
\newblock {Holographic Complexity Equals Bulk Action?}
\newblock {\em Phys. Rev. Lett.}, 116(19):191301, 2016.

\bibitem{Brown:2015lvg}
Adam~R. Brown, Daniel~A. Roberts, Leonard Susskind, Brian Swingle, and Ying
  Zhao.
\newblock {Complexity, action, and black holes}.
\newblock {\em Phys. Rev. D}, 93(8):086006, 2016.

\bibitem{Alishahiha:2015rta}
Mohsen Alishahiha.
\newblock {Holographic Complexity}.
\newblock {\em Phys. Rev. D}, 92(12):126009, 2015.

\bibitem{LloydBound}
S.~Lloyd.
\newblock {Ultimate physical limits to computation}.
\newblock {\em Nature}, 406:1047–1054, 2000.

\bibitem{Lehner:2016vdi}
Luis Lehner, Robert~C. Myers, Eric Poisson, and Rafael~D. Sorkin.
\newblock {Gravitational action with null boundaries}.
\newblock {\em Phys. Rev. D}, 94(8):084046, 2016.

\bibitem{Carmi:2017jqz}
Dean Carmi, Shira Chapman, Hugo Marrochio, Robert~C. Myers, and Sotaro
  Sugishita.
\newblock {On the Time Dependence of Holographic Complexity}.
\newblock {\em JHEP}, 11:188, 2017.

\bibitem{Cottrell:2017ayj}
William Cottrell and Miguel Montero.
\newblock {Complexity is simple!}
\newblock {\em JHEP}, 02:039, 2018.

\bibitem{Couch:2017yil}
Josiah Couch, Stefan Eccles, Willy Fischler, and Ming-Lei Xiao.
\newblock {Holographic complexity and noncommutative gauge theory}.
\newblock {\em JHEP}, 03:108, 2018.

\bibitem{Swingle:2017zcd}
Brian Swingle and Yixu Wang.
\newblock {Holographic Complexity of Einstein-Maxwell-Dilaton Gravity}.
\newblock {\em JHEP}, 09:106, 2018.

\bibitem{Alishahiha:2018tep}
Mohsen Alishahiha, Amin Faraji~Astaneh, M.~Reza Mohammadi~Mozaffar, and Ali
  Mollabashi.
\newblock {Complexity Growth with Lifshitz Scaling and Hyperscaling Violation}.
\newblock {\em JHEP}, 07:042, 2018.

\bibitem{Alishahiha:2017hwg}
Mohsen Alishahiha, Amin Faraji~Astaneh, Ali Naseh, and Mohammad~Hassan
  Vahidinia.
\newblock {On complexity for F(R) and critical gravity}.
\newblock {\em JHEP}, 05:009, 2017.

\bibitem{Mahapatra:2018gig}
Subhash Mahapatra and Pratim Roy.
\newblock {On the time dependence of holographic complexity in a dynamical
  Einstein-dilaton model}.
\newblock {\em JHEP}, 11:138, 2018.

\bibitem{Chen:2020qty}
Deyou Chen and Jie Jiang.
\newblock {Investigating the complexity-equals-action conjecture in regular
  magnetic black holes}.
\newblock {\em Class. Quant. Grav.}, 37(13):135003, 2020.

\bibitem{Babaei-Aghbolagh:2020vsz}
H.~Babaei-Aghbolagh, Komeil Babaei~Velni, Davood~Mahdavian Yekta, and
  H.~Mohammadzadeh.
\newblock {Holographic complexity for black branes with momentum relaxation}.
\newblock 9 2020.

\bibitem{Yang:2020tna}
Run-Qiu Yang, Yu-Sen An, Chao Niu, Cheng-Yong Zhang, and Keun-Young Kim.
\newblock {What kind of ''complexity'' is dual to holographic complexity?}
\newblock 11 2020.

\bibitem{Nielsen}
M.~Gu A. C.~Doherty M.~A.~Nielsen, M. R.~Dowling.
\newblock {Quantum computation as geometry}.
\newblock {\em Science}, 2006.

\bibitem{Doroudiani:2019llj}
Mehregan Doroudiani, Ali Naseh, and Reza Pirmoradian.
\newblock {Complexity for Charged Thermofield Double States}.
\newblock {\em JHEP}, 01:120, 2020.

\bibitem{HosseiniMansoori:2018gdu}
Seyed~Ali Hosseini~Mansoori, Viktor Jahnke, Mohammad~M. Qaemmaqami, and
  Yaithd~D. Olivas.
\newblock {Holographic complexity of anisotropic black branes}.
\newblock {\em Phys. Rev. D}, 100(4):046014, 2019.

\bibitem{Mateos:2011ix}
David Mateos and Diego Trancanelli.
\newblock {The anisotropic N=4 super Yang-Mills plasma and its instabilities}.
\newblock {\em Phys. Rev. Lett.}, 107:101601, 2011.

\bibitem{Mateos:2011tv}
David Mateos and Diego Trancanelli.
\newblock {Thermodynamics and Instabilities of a Strongly Coupled Anisotropic
  Plasma}.
\newblock {\em JHEP}, 07:054, 2011.

\bibitem{DHoker:2009mmn}
Eric D'Hoker and Per Kraus.
\newblock {Magnetic Brane Solutions in AdS}.
\newblock {\em JHEP}, 10:088, 2009.

\bibitem{Henningson:1998gx}
M.~Henningson and K.~Skenderis.
\newblock {The Holographic Weyl anomaly}.
\newblock {\em JHEP}, 07:023, 1998.

\bibitem{deHaro:2000vlm}
Sebastian de~Haro, Sergey~N. Solodukhin, and Kostas Skenderis.
\newblock {Holographic reconstruction of space-time and renormalization in the
  AdS / CFT correspondence}.
\newblock {\em Commun. Math. Phys.}, 217:595--622, 2001.

\bibitem{Papadimitriou:2005ii}
Ioannis Papadimitriou and Kostas Skenderis.
\newblock {Thermodynamics of asymptotically locally AdS spacetimes}.
\newblock {\em JHEP}, 08:004, 2005.

\bibitem{Bianchi:2001de}
Massimo Bianchi, Daniel~Z. Freedman, and Kostas Skenderis.
\newblock {How to go with an RG flow}.
\newblock {\em JHEP}, 08:041, 2001.

\bibitem{Martinez-y-Romero:2017awl}
Rodolfo~P. Martinez-y Romero, Leonardo Patino, and Tiber Ramirez-Urrutia.
\newblock {Increase of the Energy Necessary to Probe Ultraviolet Theories Due
  to the Presence of a Strong Magnetic Field}.
\newblock {\em JHEP}, 11:104, 2017.

\bibitem{Note1}
Our curvature convention for both models is ${R^{\alpha }}_{\beta \mu \nu
  }=\partial _{\mu }{\Gamma ^{\alpha }}_{\beta \nu }+{\Gamma ^{\alpha }}_{\mu
  \sigma }{\Gamma ^{\sigma }}_{\beta \nu }-(\mu \leftrightarrow \nu )$ and
  $R_{\mu \nu }={R^{\sigma }}_{\mu \sigma \nu }$.

\bibitem{Note2}
Taking $L=1$ implies that $G_{5}=\pi /2N_{c}^{2}$ with $N_{c}$ the number of
  color degrees of freedom in the dual theory.

\bibitem{Skenderis:2002wp}
Kostas Skenderis.
\newblock {Lecture notes on holographic renormalization}.
\newblock {\em Class. Quant. Grav.}, 19:5849--5876, 2002.

\bibitem{Bianchi:2001kw}
Massimo Bianchi, Daniel~Z. Freedman, and Kostas Skenderis.
\newblock {Holographic renormalization}.
\newblock {\em Nucl. Phys. B}, 631:159--194, 2002.

\bibitem{Ecker:2017fyh}
Christian Ecker, Carlos Hoyos, Niko Jokela, David Rodr\'\i{}guez~Fern\'andez,
  and Aleksi Vuorinen.
\newblock {Stiff phases in strongly coupled gauge theories with holographic
  duals}.
\newblock {\em JHEP}, 11:031, 2017.

\bibitem{Cvetic:1999xp}
Mirjam Cvetic, M.~J. Duff, P.~Hoxha, James~T. Liu, Hong Lu, J.~X. Lu,
  R.~Martinez-Acosta, C.~N. Pope, H.~Sati, and Tuan~A. Tran.
\newblock {Embedding AdS black holes in ten-dimensions and eleven-dimensions}.
\newblock {\em Nucl. Phys.}, B558:96--126, 1999.

\bibitem{Elinos:2021bmx}
Uriel Elinos and Leonardo Pati\~no.
\newblock {Fundamental Landau Levels in a Strongly Couple Plasma}.
\newblock 4 2021.

\bibitem{Avila:2020ved}
Daniel \'Avila and Leonardo Pati\~no.
\newblock {Melting holographic mesons by cooling a magnetized quark gluon
  plasma}.
\newblock {\em JHEP}, 06:010, 2020.

\bibitem{Arean:2016het}
Daniel Are\'an, Leopoldo~A. Pando~Zayas, Leonardo Pati\~no, and Mario
  Villasante.
\newblock {Velocity Statistics in Holographic Fluids: Magnetized Quark-Gluon
  Plasma and Superfluid Flow}.
\newblock {\em JHEP}, 10:158, 2016.

\bibitem{Avila:2018sqf}
Daniel Avila, Viktor Jahnke, and Leonardo Patiño.
\newblock {Chaos, Diffusivity, and Spreading of Entanglement in Magnetic
  Branes, and the Strengthening of the Internal Interaction}.
\newblock {\em JHEP}, 09:131, 2018.

\bibitem{Fuini:2015hba}
John~F. Fuini and Laurence~G. Yaffe.
\newblock {Far-from-equilibrium dynamics of a strongly coupled non-Abelian
  plasma with non-zero charge density or external magnetic field}.
\newblock {\em JHEP}, 07:116, 2015.

\bibitem{Endrodi:2018ikq}
Gergely Endrodi, Matthias Kaminski, Andreas Schafer, Jackson Wu, and Laurence
  Yaffe.
\newblock {Universal Magnetoresponse in QCD and $\mathcal{N}=4$ SYM}.
\newblock {\em JHEP}, 09:070, 2018.

\bibitem{Note3}
Of course, the MT and DK models satisfy this conditions for any anisotropy and
  magnetic field.

\bibitem{Note4}
Note that this means that $\tau =2t_{0}$, with $\tau $ the time variable
  relevant for the evolution of the TFD state mentioned in the introduction.

\bibitem{Carmi:2016wjl}
Dean Carmi, Robert~C. Myers, and Pratik Rath.
\newblock {Comments on Holographic Complexity}.
\newblock {\em JHEP}, 03:118, 2017.

\bibitem{Akhavan:2019zax}
Amin Akhavan and Farzad Omidi.
\newblock {On the Role of Counterterms in Holographic Complexity}.
\newblock {\em JHEP}, 11:054, 2019.

\bibitem{Chapman:2016hwi}
Shira Chapman, Hugo Marrochio, and Robert~C. Myers.
\newblock {Complexity of Formation in Holography}.
\newblock {\em JHEP}, 01:062, 2017.

\bibitem{Note5}
This term is crucial when the spacetime under consideration is not stationary,
  such as Vaidya spacetime \cite {Chapman:2018dem,Chapman:2018lsv}.

\bibitem{Reynolds:2016rvl}
Alan Reynolds and Simon~F. Ross.
\newblock {Divergences in Holographic Complexity}.
\newblock {\em Class. Quant. Grav.}, 34(10):105004, 2017.

\bibitem{Omidi:2020oit}
Farzad Omidi.
\newblock {Regularizations of Action-Complexity for a Pure BTZ Black Hole
  Microstate}.
\newblock {\em JHEP}, 07:020, 2020.

\bibitem{Note6}
It is also necessary to impose that $\protect \mathcal {F}(r_{h})=0$ by means
  of (A1).

\bibitem{Chapman:2018dem}
Shira Chapman, Hugo Marrochio, and Robert~C. Myers.
\newblock {Holographic complexity in Vaidya spacetimes. Part I}.
\newblock {\em JHEP}, 06:046, 2018.

\bibitem{Chapman:2018lsv}
Shira Chapman, Hugo Marrochio, and Robert~C. Myers.
\newblock {Holographic complexity in Vaidya spacetimes. Part II}.
\newblock {\em JHEP}, 06:114, 2018.

\end{thebibliography}
\bibliographystyle{unsrt}
\end{document}